\documentclass[12pt,a4paper]{article}
\usepackage{amsfonts}
\usepackage[dvips]{graphicx}
\usepackage{latexsym,amsmath,amssymb,graphics,stmaryrd}
\usepackage{cite}
\usepackage{array}
\usepackage{subfigure}

\usepackage{multirow}
\usepackage{epsfig}
\usepackage[dvips]{graphicx}
\usepackage[dvips]{color}
\usepackage{pstricks}
\usepackage{pst-node}
\usepackage{pst-plot}
\usepackage{dsfont}
\usepackage{amsthm}
\usepackage{graphics}
\usepackage{subfigure}
\usepackage{fancyhdr}
\usepackage[dvips]{color}
\usepackage{feynmp}

\fancypagestyle{plain}{
\fancyhf{}
\newcommand{\fpage}{\iffloatpage{}{\thepage}}
\fancyfoot[C]{\fpage}

}

\newcommand{\col}{~,}
\newcommand{\pnt}{~.}

\newcommand{\de}{\operatorname{d}\!}

\newcommand{\e}{\operatorname{e}}

\newcommand{\dchi}[1]{\boldsymbol{\chi}(#1)}

\newlength{\neglength}
\newlength{\diameter}
    
\newcommand{\svertex}[2]{%
\fmfiequ{#1}{point length(#2)/2 of #2}
}

\newcommand{\vvertex}[3]{%
\fmfipath{px}
\fmfiequ{px}{(0,ypart(#2))..(100,ypart(#2))}
\fmfiequ{#1}{point xpart(#3 intersectiontimes px) of #3}
}

\newcommand{\beq}{\begin{equation}}
\newcommand{\eeq}{\end{equation}}
\newcommand{\bea}{\begin{eqnarray}}
\newcommand{\eea}{\end{eqnarray}}
\newcommand{\ena}{\end{eqnarray}}

\newcommand{\id}{\mathrm{d}}

\renewcommand{\a}{\alpha}
\renewcommand{\b}{\beta}

\newcommand{\g}{\gamma}

\newcommand{\z}{\zeta}

\renewcommand{\l}{\lambda}

\newcommand{\p}{\pi}

\newcommand{\plainwrap}[4]{%
\fmfipath{pi[]}
\fmfiset{pi1}{vloc(__#1) ..controls (-0.175w,ypart(vloc(__#1))) and (-0.175w,-0.15w) .. (xpart(vloc(__#2)),-0.15w)}
\fmfiset{pi2}{(xpart(vloc(__#2)),-0.15w) ..(xpart(vloc(__#3)),-0.15w)}

\fmfiset{pi3}{(xpart(vloc(__#3)),-0.15w) ..controls (1.175w,-0.15w) and (1.175w,ypart(vloc(__#4))) .. vloc(__#4)}
\fmfi{plain}{pi1 ..pi2 ..pi3}
}
\newcommand{\wigglywrap}[4]{%
\fmfipath{pi[]}
\fmfiset{pi1}{#1 ..controls (-0.175w,ypart(#1)) and (-0.175w,-0.15w) .. (xpart(vloc(__#2)),-0.15w)}
\fmfiset{pi2}{(xpart(vloc(__#2)),-0.15w) ..(xpart(vloc(__#3)),-0.15w)}

\fmfiset{pi3}{(xpart(vloc(__#3)),-0.15w) ..controls (1.175w,-0.15w) and (1.175w,ypart(#4)) .. #4}
\fmfi{wiggly}{pi1 ..pi2 ..pi3}
}

\DeclareMathOperator{\tr}{tr}

\numberwithin{equation}{section}
\newlength{\eqoff}
\newlength{\eqofftwo}
\newlength{\unit}
\psset{xunit=\unit,yunit=\unit,runit=\unit}
\newlength{\linew}
\psset{linewidth=\linew}

\begin{document}
\begin{fmffile}{fullgraphs}

\fmfcmd{
wiggly_len := 2mm;
vardef wiggly expr p_arg =
 save wpp,len;
 numeric wpp,alen;
 wpp = ceiling (pixlen (p_arg, 10) / wiggly_len);
 len=length p_arg;
 for k=0 upto wpp - 1:
  point arctime k/(wpp-1)*arclength(p_arg) of p_arg of p_arg
    {direction arctime k/(wpp-1)*arclength(p_arg) of p_arg of p_arg rotated wiggly_slope} ..
  point  arctime (k+.5)/(wpp-1)*arclength(p_arg) of p_arg of p_arg
 {direction arctime (k+.5)/(wpp-1)*arclength(p_arg) of p_arg of p_arg rotated - wiggly_slope} ..
 endfor
 if cycle p_arg: cycle else: point infinity of p_arg fi
enddef;
}
\fmfcmd{%
marksize=1.5mm;
def draw_mark(expr p,a) =
  begingroup
    save t,tip,dma,dmb; pair tip,dma,dmb;
    t=arctime a of p;
    tip =marksize*unitvector direction t of p;
    dma =marksize*unitvector direction t of p rotated -45;
    dmb =marksize*unitvector direction t of p rotated 45;
    linejoin:=beveled;
    draw (-.5dma.. .5tip-- -.5dmb) shifted point t of p;
  endgroup
enddef;
style_def derplain expr p =
    save amid;
    amid=.5*arclength p;
    draw_mark(p, amid);
    draw p;
enddef;
def draw_point(expr p,a) =
  begingroup
    save t,tip,dma,dmb,dmo; pair tip,dma,dmb,dmo;
    t=arctime a of p;
    tip =marksize*unitvector direction t of p;
    dma =marksize*unitvector direction t of p rotated -45;
    dmb =marksize*unitvector direction t of p rotated 45;
    dmo =marksize*unitvector direction t of p rotated 90;
    linejoin:=beveled;
    draw (-.5dma.. .5tip-- -.5dmb) shifted point t of p withcolor 0white;
    draw (-.5dmo.. .5dmo) shifted point t of p;
  endgroup
enddef;
style_def derplainpt expr p =
    save amid;
    amid=.5*arclength p;
    draw_point(p, amid);
    draw p;
enddef;
style_def dblderplain expr p =
    save amidm;
    save amidp;
    amidm=.5*arclength p-0.5mm;
    amidp=.5*arclength p+0.5mm;
    draw_mark(p, amidm);
    draw_point(p, amidp);
    draw p;
enddef;
}

\begin{titlepage}
\begin{flushright}
IFUM-934-FT \\
\end{flushright}

\Large
\begin {center}     
{\bf
Single impurity operators at critical wrapping order 
in the  $\b$-deformed ${\cal{N}}=4$ SYM}
\end {center}

\renewcommand{\thefootnote}{\fnsymbol{footnote}}

\large
\vspace{1cm}
\centerline{F.\ Fiamberti ${}^{a,b}$, A.\ Santambrogio ${}^b$, 
C.\ Sieg ${}^c$, D.\ Zanon ${}^{a,b}$
\footnote[1]{\noindent \tt
francesco.fiamberti@mi.infn.it \\
\hspace*{6.3mm}alberto.santambrogio@mi.infn.it \\ 
\hspace*{6.3mm}csieg@nbi.dk \\ 
\hspace*{6.3mm}daniela.zanon@mi.infn.it}}
\vspace{4ex}
\normalsize
\begin{center}
\emph{$^a$  Dipartimento di Fisica, Universit\`a degli Studi di Milano, \\
Via Celoria 16, 20133 Milano, Italy}\\
\vspace{0.2cm}
\emph{$^b$ INFN--Sezione di Milano,\\
Via Celoria 16, 20133 Milano, Italy}\\
\vspace{0.2cm}
\emph{$^c$ The Niels Bohr International Academy,\\
The Niels Bohr Institute,\\
Blegdamsvej 17, DK-2100 Copenhagen, Denmark\\
}
\end{center}
\vspace{0.5cm}
\rm
\abstract
\normalsize 
\noindent We study  the spectrum of one single magnon  in the superconformal $\b$-deformed ${\cal{N}}=4$ SYM theory in the planar limit. We compute the anomalous dimensions of one-impurity operators ${\cal{O}}_{1,L}= \tr(\phi Z^{L-1})$, including wrapping contributions at their critical order $L$. 
\vfill
\end{titlepage} 

\section{Introduction}
In some recent papers \cite{us,uslong,betadef} we have studied finite size effects in the anomalous dimensions of gauge invariant composite operators. More precisely we have addressed this issue  performing the analysis in two special settings, i.e. in the   planar limit of the superconformal $\mathcal{N}=4$ SYM theory and in the exactly marginal deformation of ${\cal N}=4$ SYM theory preserving ${\cal N}=1$ supersymmetry.

The interest in this kind of calculations stems from the quest of an ever deeper  understanding of the
AdS/CFT correspondence \cite{maldacena:1998re} and  the comparison of the spectra of the gauge theory and the string theory. On the gauge theory side important progress has been achieved thanks to the realization that some sectors of the theory can be described in terms of quantum spin chains  for which in many cases an Hamiltonian, an asymptotic Bethe ansatz \cite{Minahan:2002ve,Beisert:2003tq,Beisert:2003ys,Staudacher:2004tk,Beisert:2004hm} and even an exact, factorized S-matrix corrected by a dressing phase~\cite{Arutyunov:2004vx,Hernandez:2006tk,Beisert:2006ib,Beisert:2006ez} are available. These results were primarily obtained for the anomalous dimensions of operators of infinite length whose analysis is considerably simplified. 

When finite size effects become important new interactions need to be taken into account.
On the string theory side recent papers have studied finite size contributions in the spectrum of magnons~\cite{SchaferNameki:2006gk,SchaferNameki:2006ey,Arutyunov:2006gs,Minahan:2008re,Heller:2008at,Ramadanovic:2008qd,Hatsuda:2008gd,Gromov:2008ie,Sax:2008in,Hatsuda:2008na}. On the field theory side wrapping interactions~\cite{Beisert:2004hm} have been analyzed \cite{Sieg:2005kd} in the $\mathcal{N}=4$ SYM theory.   The anomalous dimension of the composite operator  $ \tr(\phi[Z,\phi]Z)$ has been computed  at four loops in \cite{us,uslong,Bajnok:2008bm,Velizhanin:2008jd}.

In \cite{betadef} we have considered finite size effects in a less symmetric setting.  More precisely we have focused on  the $\b$-deformed ${\cal{N}}=4$ SYM theory obtained from the original ${\cal N}=4$ theory modifying the superpotential via the substitution
\beq
ig\,\tr\left(\phi\,\psi\,Z -  \phi\,Z\,\psi\right)~\longrightarrow ~ih\,\tr\left(\e^{i\pi\b} \phi\,\psi\,Z - \e^{-i\pi\b} \phi\,Z\,\psi\right)
\label{deformation}
\eeq
with  $h$ and $\b$  complex constants.
In \cite{Leigh:1995ep} it was argued that this
theory becomes conformally invariant if the constants $h$ and $\b$ satisfy one condition.  Indeed in \cite{Mauri:2005pa} it has been shown that for a real deformation parameter   $\b$ this deformed ${\cal N}=1$ theory becomes superconformal, in the planar limit to all perturbative orders, if 
\beq
h\bar{h}=g^2\pnt
\label{betareal}
\eeq
The AdS/CFT correspondence predicts that this  deformed theory  be equivalent to the Lunin-Maldacena string theory background~\cite{Lunin:2005jy}. The existence of integrable structures in the deformed string background has been analyzed in~\cite{Frolov:2005ty,Berenstein:2004ys}. 

A new feature of the deformed theory as compared to the situation in ${\cal N}=4$ SYM is given by the fact that  one-impurity states are not protected by supersymmetry. In \cite{betadef} we have computed the anomalous dimensions of one-impurity and two-impurity operators up to four-loop order and made some  partial calculations and conjectures for
the simplest single-impurity operator ${\cal{O}}_{1,L}= \tr(\phi Z^{L-1})$ at higher order $L$ in perturbation theory. 

This paper is the natural extension of the work presented in \cite{betadef}. Here we want to compute  in the planar limit the anomalous dimensions of the one-impurity operators ${\cal{O}}_{1,L}= \tr(\phi Z^{L-1})$, including wrapping contributions at their critical order $L$.

The anomalous dimension of a composite operator $\mathcal{O}$ can be obtained from the divergent diagrams that contribute to its one-point function. In dimensional regularization, if the operator is renormalized multiplicatively, it  is simply given by
\begin{equation}
\gamma(\mathcal{O})=\lim_{\varepsilon\rightarrow0}\left[\varepsilon g\frac{\mathrm{d} }{\mathrm{d} g}\log\mathcal{Z}_{\mathcal{O}}(g,\varepsilon)\right]\col
\label{anomdim}
\end{equation}
where
\begin{equation}
\mathcal{O}_{\mathrm{ren}}=\mathcal{Z}_\mathcal{O}\mathcal{O}_{\mathrm{bare}}
\label{opren}
\end{equation}
and $\varepsilon$ is the dimensional regularization parameter. Thus we want to compute up to $L$ loops all the divergent contributions to the one-point function of the operator ${\cal{O}}_{1,L}$ and isolate the $1/\varepsilon$  poles.
At first sight this program looks very ambitious and complicated. In fact  the use in conjunction of integrability properties and of superspace techniques proved  so powerful that we were able to reduce and structure the whole calculation into a manageable form. The computation is organized through the sequence of the following steps:\\
First we obtain the contributions to the anomalous dimension which do not contain wrapping interactions, up to perturbative order $L$. This we achieve by starting from the knowledge of the anomalous dimensions of  long single-impurity states of the SU(2) sector.     If the corresponding operator $\mathcal{O}_\text{as}$ has such a length that wrapping interactions do not contribute, the anomalous dimension  at a given perturbative order  can be obtained from the all-loop result~\cite{Mauri:2005pa}
\begin{equation}
\gamma(\mathcal{O}_\text{as})=-1+\sqrt{1+4\lambda\Big\vert q-\frac{1}{q}\Big\vert^2}=-1+\sqrt{1+16\lambda\sin^2(\pi\beta)}
\col
\label{gammaas}
\end{equation}
where $\lambda=\frac{g^2N}{16\pi^2}$. This result is correct only in the asymptotic regime. \\
 Thus in order to obtain the anomalous dimensions for ${\cal{O}}_{1,L}$ up to order $L$, we need to subtract the range-$(L+1)$ contributions from the $L$-order expansion of (\ref{gammaas}) and finally compute explicitly the wrapping diagrams at the critical order $L$.\\
 The number of wrapping graphs that one has to consider at $L$ loops becomes so large that in \cite{betadef} we thought the calculation could not be performed exactly. In fact a judicious 
analysis in terms of superfields and supergraphs allows to discover a huge number of cancellations such that in the end only few restricted classes of supergraphs are shown to be relevant.
In the next section we consider these superspace Feynman diagram calculations, showing how the cancellations occur and performing the $D$-algebra on the relevant contributions. In section 3 we study the integrals. In section 4 we comment on our results. The strategy to explicitly compute the required integrals can be found in the appendix.

\section{The power of superspace}

We want to compute the anomalous dimension of the
composite operator ${\cal{O}}_{1,L}= \tr(\phi Z^{L-1})$, including wrapping contributions up to  $L$ loops  in the planar limit following the strategy introduced in \cite{us,uslong,betadef}.  While doing similar calculations at lower loop order  we learned that superspace techniques are very efficient for organizing the work and moreover we  found many unexpected cancellations to occur. In this section we will review briefly the general rules of the ${\cal N}=1$ superspace formalism, primarily to set our notations, and we will explain the reasons that lead to the many cancellations we had noticed in our previous works. It has been the realization of these general great simplifications that encouraged us in undertaking this project.
 
The action of $\mathcal{N}=1$ $\beta$-deformed SYM is described in terms of
 one real vector superfield $V$ and three
chiral superfields $\phi,\psi,Z$ that we denote collectively by $\phi^i$. Following the notations and conventions of \cite{Gates:1983nr} it is given by

\begin{equation}\label{action}
\begin{aligned}
S &= \int \id^4 x\, \id^4 \theta \, \tr \left(\e^{-gV} \bar \phi_i \e^{gV}
\phi^i\right) + \frac 1{2g^2} \int \id^4 x\, \id^2 \theta
\,\tr \left(W^\alpha W_\alpha\right)\\
&\phantom{{}={}}
+ih  \int \id^4 x\, \id^2 \theta \tr\left(\e^{i\pi\b} \phi_1\,\phi_2\,\phi_3 - \e^{-i\pi\b} \phi_1\,\phi_3\,\phi_2\right)
 + \text{h.c.}
\col
\end{aligned}
\end{equation}
where $W_\alpha = \bar D^2 \left(\e^{-gV} D_\alpha\,\e^{gV}\right)$, and
$V=V^aT^a$, $\phi^i=\phi_i^aT^a$, {\small $i=1,2,3$}, $T^a$ being
matrices satisfying the $SU(N)$ algebra
\begin{equation}
[T_a, T_b] = i f_{abc} T_c
\end{equation}
and normalized as
\begin{equation}
\tr(T_a T_b) =\delta_{ab}
\pnt
\end{equation}
In order to compute Feynman diagrams we need propagators and vertices that can be easily obtained from the 
action (\ref{action}).
The superfield propagators are given in momentum space by
\begin{equation}
\langle V^a V^b\rangle= - \frac{\delta^{ab}}{p^2}\col\qquad\qquad
\langle\phi^a_i \bar\phi^b_j\rangle=\delta_{ij} \frac{\delta^{ab}}{p^2}
\pnt
\label{propagators}
\end{equation}
The vertices  that we need are
\begin{equation}\label{vertices}
\begin{aligned}
V_1&=\ g f_{abc}\delta^{ij} \bar\phi^a_i V^b \phi^c_j \col\qquad\qquad\qquad
V_2=\frac{g^2}{2}  \delta^{ij} f_{adm} f_{bcm} V^aV^b \bar\phi^c_i
\phi^d_j\col\\
V_3&= - h f_{abc} (\e^{i\pi\b} \phi_1^a\,\phi_2^b\,\phi_3^c - \e^{-i\pi\b} \phi_1^a\,\phi_3^b\,\phi_2^c)
\col\\
\bar{V}_3&= - \bar{h} f_{abc} (\e^{-i\pi\b}\bar\phi_1^a \bar\phi_2^b \bar\phi_3^c-\e^{i\pi\b} \bar\phi_1^a \bar\phi_3^b \bar\phi_2^c)\col
\end{aligned}
\end{equation}
with additional $\bar D^2$, $D^2$ factors for each chiral, antichiral line
respectively. It is easy to realize that since our operator ${\cal{O}}_{1,L}$ has  length $L$,  vertices containing three or more vector superfields $V$ never enter the $L$-loop calculation.

As anticipated in the introduction essentially all diagrams without wrapping interactions need not be examined since their contribution to the $L$-loop anomalous dimension of ${\cal{O}}_{1,L}$ is simply obtained from the $L$-order expansion of the asymptotic result in (\ref{gammaas})
\beq
\g^{as}_L= \alpha_L\,\l^L \sin^{2L}(\p\b)\col\qquad\alpha_L=-(-8)^L \frac{(2L-3)!!}{L!}
\pnt
\label{asym}
\eeq
In fact the above result is valid for single-impurity states of length greater than $L$. Thus we need correct it by subtracting the range-$(L+1)$ contributions. Here is where our Feynman diagram computation really starts. It is organized in two separate steps:\\
a) subtraction of the range $L+1$ diagrams\\
b) computation of all the $L$-loop wrapping diagrams.

According to the general procedure  one has to consider a given supergraph and complete the D-algebra in order to reduce it to a standard divergent graph. All graphs that give rise to finite integrals are immediately discarded since they are not relevant for the computation of the anomalous dimension of the
composite operator.
In  \cite{uslong} we have shown that in order to produce divergent contributions  the D-algebra  must be performed in such a way that no spinor derivative is moved out  onto the external lines, except for derivatives on scalar propagators that do not belong to any loop. The following analysis makes always use of this very strong result. 

\begin{figure}[t]
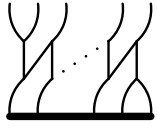

\unitlength=0.75mm
\settoheight{\eqoff}{$\times$}%
\setlength{\eqoff}{0.5\eqoff}%
\addtolength{\eqoff}{-12.5\unitlength}%
\settoheight{\eqofftwo}{$\times$}%
\setlength{\eqofftwo}{0.5\eqofftwo}%
\addtolength{\eqofftwo}{-7.5\unitlength}%
\centering
\raisebox{\eqoff}{%
\fmfframe(3,1)(1,4){%
\begin{fmfchar*}(25,20)
\fmftop{v1}
\fmfbottom{v5}
\fmfforce{(0w,h)}{v1}
\fmfforce{(0w,0)}{v5}
\fmffixed{(0.2w,0)}{v1,v2}
\fmffixed{(0.2w,0)}{v2,v3}
\fmffixed{(0.2w,0)}{v3,w1}
\fmffixed{(0.2w,0)}{w1,v4}
\fmffixed{(0.2w,0)}{v4,v9}
\fmffixed{(0.2w,0)}{v5,v6}
\fmffixed{(0.2w,0)}{v6,v7}
\fmffixed{(0.2w,0)}{v7,w2}
\fmffixed{(0.2w,0)}{w2,v8}
\fmffixed{(0.2w,0)}{v8,v10}
\fmffixed{(0,whatever)}{vc1,vc2}
\fmffixed{(0,whatever)}{vc3,vc4}
\fmffixed{(0,whatever)}{vc5,vc6}
\fmffixed{(0,whatever)}{vc7,vc8}
\fmf{plain,tension=0.25,right=0.25}{v1,vc1}
\fmf{plain,tension=0.25,left=0.25}{v2,vc1}
\fmf{plain,left=0.25}{v5,vc2}
\fmf{plain,tension=1,left=0.25}{v3,vc3}
\fmf{phantom,tension=1,left=0.25}{w1,wc1}
\fmf{plain,tension=1,left=0.25}{v4,vc5}
\fmf{plain,tension=1,left=0.25}{v9,vc7}
\fmf{plain,left=0.25}{w2,vc6}
\fmf{plain,tension=0.25,left=0.25}{v8,vc8}
\fmf{plain,tension=0.25,right=0.25}{v10,vc8}
\fmf{plain,left=0.25}{v6,vc4}
\fmf{phantom,left=0.25}{v7,wc2}
\fmf{plain,tension=0.5}{vc1,vc2}
\fmf{phantom,tension=0.5}{wc1,wc2}
\fmf{plain,tension=0.5}{vc2,vc3}
\fmf{phantom,tension=0.5}{wc2,vc5}
\fmf{plain,tension=0.5}{vc3,vc4}
\fmf{phantom,tension=0.5}{vc4,wc1}
\fmf{plain,tension=0.5}{vc5,vc6}
\fmf{plain,tension=0.5}{vc6,vc7}
\fmf{plain,tension=0.5}{vc7,vc8}
\fmf{plain,tension=0.5,right=0,width=1mm}{v5,v10}
\fmffreeze
\fmf{dots,tension=0.5}{vc4,vc5}
\fmfposition
\end{fmfchar*}}}
\caption{$L$-loop, range-$(L+1)$ diagram $S_L$}
\label{diagram-SL}
\end{figure}

\begin{figure}[h]
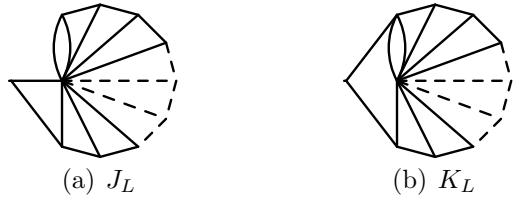

\unitlength=0.75mm
\settoheight{\eqoff}{$\times$}%
\setlength{\eqoff}{0.5\eqoff}%
\addtolength{\eqoff}{-12.5\unitlength}%
\settoheight{\eqofftwo}{$\times$}%
\setlength{\eqofftwo}{0.5\eqofftwo}%
\addtolength{\eqofftwo}{-7.5\unitlength}%
\centering
\subfigure[$J_L$]{
\label{JL}
\raisebox{\eqoff}{%
\fmfframe(0,0)(0,0){%
\begin{fmfchar*}(30,30)
  \fmfleft{in}
  \fmfright{out}
  \fmf{plain,tension=1}{in,vi}
  \fmf{phantom,tension=1}{out,v7}
\fmfpoly{phantom}{v12,v11,v10,v9,v8,v7,v6,v5,v4,v3,v2,v1}
\fmffixed{(0.9w,0)}{v1,v7}
\fmffixed{(0.3w,0)}{vi,v0}
\fmffixed{(0,whatever)}{v0,v3}
\fmf{plain}{v3,v4}
\fmf{plain}{v4,v5}
\fmf{plain}{v5,v6}
\fmf{dashes}{v6,v7}
\fmf{dashes}{v7,v8}
\fmf{dashes}{v8,v9}
\fmf{plain}{v9,v10}
\fmf{plain}{v10,v11}
\fmf{phantom}{v3,vi}
\fmf{plain}{vi,v11}
\fmf{phantom}{v0,v3}
\fmf{plain}{v0,v4}
\fmf{plain}{v0,v5}
\fmf{plain}{v0,v6}
\fmf{dashes}{v0,v7}
\fmf{dashes}{v0,v8}
\fmf{plain}{v0,v9}
\fmf{plain}{v0,v10}
\fmf{plain}{v0,v11}
\fmffreeze
\fmfposition
\fmf{plain}{vi,v0}
\fmf{plain,left=0.25}{v0,v3}
\fmf{plain,left=0.25}{v3,v0}
\fmfipair{w[]}
\fmfiequ{w1}{(xpart(vloc(__v3)),ypart(vloc(__v3)))}
\fmfiequ{w2}{(xpart(vloc(__v4)),ypart(vloc(__v4)))}
\fmfiequ{w3}{(xpart(vloc(__v5)),ypart(vloc(__v5)))}
\fmfiequ{w4}{(xpart(vloc(__v6)),ypart(vloc(__v6)))}
\fmfiequ{w5}{(xpart(vloc(__v7)),ypart(vloc(__v7)))}
\fmfiequ{w6}{(xpart(vloc(__v8)),ypart(vloc(__v8)))}
\fmfiequ{w7}{(xpart(vloc(__v9)),ypart(vloc(__v9)))}
\end{fmfchar*}}
}
}
\qquad\qquad
\subfigure[$K_L$]{
\label{KL}
\raisebox{\eqoff}{%
\fmfframe(0,0)(0,0){%
\begin{fmfchar*}(30,30)
  \fmfleft{in}
  \fmfright{out}
  \fmf{plain,tension=1}{in,vi}
  \fmf{phantom,tension=1}{out,v7}
\fmfpoly{phantom}{v12,v11,v10,v9,v8,v7,v6,v5,v4,v3,v2,v1}
\fmffixed{(0.9w,0)}{v1,v7}
\fmffixed{(0.3w,0)}{vi,v0}
\fmffixed{(0,whatever)}{v0,v3}
\fmf{plain}{v3,v4}
\fmf{plain}{v4,v5}
\fmf{plain}{v5,v6}
\fmf{dashes}{v6,v7}
\fmf{dashes}{v7,v8}
\fmf{dashes}{v8,v9}
\fmf{plain}{v9,v10}
\fmf{plain}{v10,v11}
\fmf{plain}{v3,vi}
\fmf{plain}{vi,v11}
\fmf{phantom}{v0,v3}
\fmf{plain}{v0,v4}
\fmf{plain}{v0,v5}
\fmf{plain}{v0,v6}
\fmf{dashes}{v0,v7}
\fmf{dashes}{v0,v8}
\fmf{plain}{v0,v9}
\fmf{plain}{v0,v10}
\fmf{plain}{v0,v11}
\fmffreeze
\fmfposition
\fmf{plain,left=0.25}{v0,v3}
\fmf{plain,left=0.25}{v3,v0}
\fmfipair{w[]}
\fmfiequ{w1}{(xpart(vloc(__v3)),ypart(vloc(__v3)))}
\fmfiequ{w2}{(xpart(vloc(__v4)),ypart(vloc(__v4)))}
\fmfiequ{w3}{(xpart(vloc(__v5)),ypart(vloc(__v5)))}
\fmfiequ{w4}{(xpart(vloc(__v6)),ypart(vloc(__v6)))}
\fmfiequ{w5}{(xpart(vloc(__v7)),ypart(vloc(__v7)))}
\fmfiequ{w6}{(xpart(vloc(__v8)),ypart(vloc(__v8)))}
\fmfiequ{w7}{(xpart(vloc(__v9)),ypart(vloc(__v9)))}
\end{fmfchar*}}
}
}
\caption{$L$-loop integrals from diagrams $S_L$ and $W_{L,0}$}
\label{integrals-SWL}
\end{figure}

Now we start considering the Feynman diagrams that we have to subtract in order to cancel the unwanted range-$(L+1)$ contributions contained in (\ref{asym}).  As in the lower order examples considered in \cite{betadef}, we have to subtract only one diagram (and its reflection) constructed entirely in terms of chiral interactions. It is shown in Fig.~\ref{diagram-SL} and denoted by $S_L$. Introducing the same notations for the chiral structures of the supergraphs as in \cite{betadef}, we find
\begin{equation}
\begin{aligned}
S_L&\rightarrow
(g^2 N)^L J_L\,[\dchi{1,2,\dots,L}+
\dchi{L,\dots,2,1}]\\
&\phantom{{}\rightarrow{}}
=(g^2 N)^L J_L\, 
(q-\bar{q})^2\left[q^{2(L-1)}+\bar{q}^{2(L-1)}
\right]\col
\end{aligned}
\end{equation}
where $J_L$ is given by the L-loop integral in Fig.~\ref{JL}.
Thus the term to be subtracted from~\eqref{asym} is
\begin{equation}
\delta\gamma_L^{\,\text{s}}=-2L\lim_{\varepsilon\rightarrow0}(\varepsilon S_L)
\pnt
\end{equation}

\begin{figure}[t]
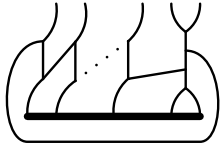

\unitlength=0.75mm
\settoheight{\eqoff}{$\times$}%
\setlength{\eqoff}{0.5\eqoff}%
\addtolength{\eqoff}{-12.5\unitlength}%
\settoheight{\eqofftwo}{$\times$}%
\setlength{\eqofftwo}{0.5\eqofftwo}%
\addtolength{\eqofftwo}{-7.5\unitlength}%
\centering
\raisebox{\eqoff}{%
\fmfframe(3,1)(1,4){%
\begin{fmfchar*}(30,20)
\fmftop{v1}
\fmfbottom{v5}
\fmfforce{(0w,h)}{v1}
\fmfforce{(0w,0)}{v5}
\fmffixed{(0.17w,0)}{v1,v2}
\fmffixed{(0.17w,0)}{v2,v3}
\fmffixed{(0.17w,0)}{v3,w1}
\fmffixed{(0.17w,0)}{w1,w3}
\fmffixed{(0.17w,0)}{w3,v4}
\fmffixed{(0.17w,0)}{v4,v9}
\fmffixed{(0.17w,0)}{v5,v6}
\fmffixed{(0.17w,0)}{v6,v7}
\fmffixed{(0.17w,0)}{v7,w2}
\fmffixed{(0.17w,0)}{w2,w4}
\fmffixed{(0.17w,0)}{w4,v8}
\fmffixed{(0.17w,0)}{v8,v10}
\fmffixed{(0,whatever)}{vc1,vc2}
\fmffixed{(0,whatever)}{vc3,vc4}
\fmffixed{(0,whatever)}{vc5,vc6}
\fmffixed{(0,whatever)}{vc7,vc8}
\fmffixed{(0,whatever)}{wc3,wc4}
\fmf{phantom,tension=0.25,right=0.25}{v1,vc1}
\fmf{plain,tension=0.25,left=0.25}{v2,vc1}
\fmf{plain,left=0.25}{v5,vc2}
\fmf{plain,tension=1,left=0.25}{v3,vc3}
\fmf{phantom,tension=1,left=0.25}{v3,wc3}
\fmf{phantom,tension=1,left=0.25}{w1,wc3}
\fmf{phantom}{wc3,wc4}
\fmf{plain,tension=0.5,right=0.25}{v4,vc7}
\fmf{plain,tension=0.5,left=0.25}{v9,vc7}
\fmf{plain,tension=0.5,left=0.25}{v8,vc8}
\fmf{plain,tension=0.5,right=0.25}{v10,vc8}
\fmf{plain,left=0.25}{v6,vc4}
\fmf{phantom,right=0.25}{v7,vc4}
\fmf{phantom}{vc4,wc3}
\fmf{plain,tension=0.5}{vc1,vc2}
\fmf{plain,tension=0.5}{vc2,vc3}
\fmf{plain,tension=0.5}{vc3,vc4}
\fmf{plain,tension=0.5}{vc7,vc8}
\fmf{plain,tension=0.5,right=0,width=1mm}{v5,v10}
\fmf{plain,tension=0.5}{vc5,vc6}
\fmffreeze
\fmf{plain,tension=0.25,left=0.25}{w2,vc6}
\fmf{phantom,tension=0.25,right=0.25}{w4,vc6}
\fmf{phantom,tension=0.25,right=0.25}{w1,wc1}
\fmf{plain,tension=0.25,left=0.25}{w3,wc1}
\fmf{plain,tension=0.5}{vc6,wc1}
\fmffreeze
\fmf{dots}{vc4,wc1}
\fmfposition
\fmfipath{p[]}
\fmfiset{p1}{vpath(__vc8,__vc7)}
\fmfipair{wz[]}
\fmfiequ{wz1}{point length(p1)/3 of p1}
\fmfiequ{wz2}{point 2*length(p1)/3 of p1}
\fmfiequ{wz3}{(xpart(vloc(__vc6)),ypart(vloc(__vc6)))}
\fmfiequ{wz4}{(xpart(vloc(__vc1)),ypart(vloc(__vc1)))}
\fmf{plain}{vc6,v12}
\fmfforce{(xpart(wz1),ypart(wz1))}{v11}
\fmfforce{(xpart(wz2),ypart(wz2))}{v12}
\plainwrap{vc1}{v5}{v10}{v11}
\end{fmfchar*}}}
\caption{$L$-loop wrapping diagram $W_{L,0}$}
\label{diagram-WL0}
\end{figure}

Now we concentrate on the wrapping diagrams. There is one wrapping graph (and its reflection) with only chiral interactions and it is depicted in Fig.~\ref{diagram-WL0}. With the identification of the first and $(L+1)$-th lines we can describe its chiral structure in terms of the deformed ones and we find
\begin{equation}
\begin{aligned}
W_{L,0}&\rightarrow(g^2 N)^L K_L\,[\dchi{1,L,L-1,\dots,2}+\dchi{L,1,2,\dots,L-1}]\\
&\phantom{{}\rightarrow{}}=(g^2 N)^L K_L\,
(q-\bar{q})^2\left[q^{2(L-1)}+\bar{q}^{2(L-1)}
\right]\pnt
\end{aligned}
\end{equation}
The integral $K_L$ is given in Fig.~\ref{KL}.

\begin{figure}[t]
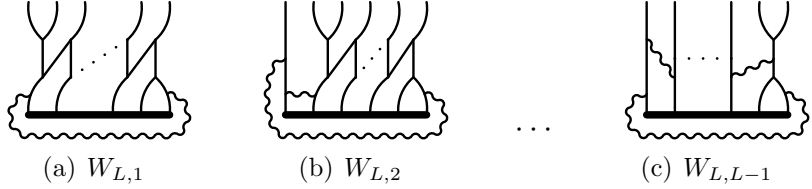

\unitlength=0.75mm
\settoheight{\eqoff}{$\times$}%
\setlength{\eqoff}{0.5\eqoff}%
\addtolength{\eqoff}{-12.5\unitlength}%
\settoheight{\eqofftwo}{$\times$}%
\setlength{\eqofftwo}{0.5\eqofftwo}%
\addtolength{\eqofftwo}{-7.5\unitlength}%
\centering
\raisebox{\eqoff}{%
\subfigure[$W_{L,1}$]
{
\fmfframe(3,1)(1,4){%
\begin{fmfchar*}(25,20)
\fmftop{v1}
\fmfbottom{v5}
\fmfforce{(0w,h)}{v1}
\fmfforce{(0w,0)}{v5}
\fmffixed{(0.2w,0)}{v1,v2}
\fmffixed{(0.2w,0)}{v2,v3}
\fmffixed{(0.2w,0)}{v3,w1}
\fmffixed{(0.2w,0)}{w1,v4}
\fmffixed{(0.2w,0)}{v4,v9}
\fmffixed{(0.2w,0)}{v5,v6}
\fmffixed{(0.2w,0)}{v6,v7}
\fmffixed{(0.2w,0)}{v7,w2}
\fmffixed{(0.2w,0)}{w2,v8}
\fmffixed{(0.2w,0)}{v8,v10}
\fmffixed{(0,whatever)}{vc1,vc2}
\fmffixed{(0,whatever)}{vc3,vc4}
\fmffixed{(0,whatever)}{vc5,vc6}
\fmffixed{(0,whatever)}{vc7,vc8}
\fmf{plain,tension=0.25,right=0.25}{v1,vc1}
\fmf{plain,tension=0.25,left=0.25}{v2,vc1}
\fmf{plain,left=0.25}{v5,vc2}
\fmf{plain,tension=1,left=0.25}{v3,vc3}
\fmf{phantom,tension=1,left=0.25}{w1,wc1}
\fmf{plain,tension=1,left=0.25}{v4,vc5}
\fmf{plain,tension=1,left=0.25}{v9,vc7}
\fmf{plain,left=0.25}{w2,vc6}
\fmf{plain,tension=0.25,left=0.25}{v8,vc8}
\fmf{plain,tension=0.25,right=0.25}{v10,vc8}
\fmf{plain,left=0.25}{v6,vc4}
\fmf{phantom,left=0.25}{v7,wc2}
\fmf{plain,tension=0.5}{vc1,vc2}
\fmf{phantom,tension=0.5}{wc1,wc2}
\fmf{plain,tension=0.5}{vc2,vc3}
\fmf{phantom,tension=0.5}{wc2,vc5}
\fmf{plain,tension=0.5}{vc3,vc4}
\fmf{phantom,tension=0.5}{vc4,wc1}
\fmf{plain,tension=0.5}{vc5,vc6}
\fmf{plain,tension=0.5}{vc6,vc7}
\fmf{plain,tension=0.5}{vc7,vc8}
\fmf{plain,tension=0.5,right=0,width=1mm}{v5,v10}
\fmffreeze
\fmf{dots,tension=0.5}{vc4,vc5}
\fmfposition
\fmfipath{p[]}
\fmfiset{p1}{vpath(__v5,__vc2)}
\fmfiset{p2}{vpath(__v10,__vc8)}
\fmfipair{wz[]}
\fmfiequ{wz2}{point length(p2)/2 of p2}
\vvertex{wz1}{wz2}{p1}
\svertex{wz2}{p2}
\wigglywrap{wz1}{v5}{v10}{wz2}
\fmfposition
\end{fmfchar*}}
}
\qquad
\subfigure[$W_{L,2}$]{
\fmfframe(3,1)(1,4){%
\begin{fmfchar*}(25,20)
\fmftop{v1}
\fmfbottom{v5}
\fmfforce{(0w,h)}{v1}
\fmfforce{(0w,0)}{v5}
\fmffixed{(0.2w,0)}{v1,v2}
\fmffixed{(0.2w,0)}{v2,v3}
\fmffixed{(0.2w,0)}{v3,w1}
\fmffixed{(0.2w,0)}{w1,v4}
\fmffixed{(0.2w,0)}{v4,v9}
\fmffixed{(0.2w,0)}{v5,v6}
\fmffixed{(0.2w,0)}{v6,v7}
\fmffixed{(0.2w,0)}{v7,w2}
\fmffixed{(0.2w,0)}{w2,v8}
\fmffixed{(0.2w,0)}{v8,v10}
\fmffixed{(0,whatever)}{vc1,vc2}
\fmffixed{(0,whatever)}{vc3,vc4}
\fmffixed{(0,whatever)}{vc5,vc6}
\fmffixed{(0,whatever)}{vc7,vc8}
\fmf{plain,tension=0.25,right=0.25}{v2,vc1}
\fmf{plain,tension=0.25,left=0.25}{v3,vc1}
\fmf{plain,left=0.25}{v6,vc2}
\fmf{plain,tension=1,left=0.25}{w1,vc3}
\fmf{phantom,tension=1,left=0.25}{v4,wc1}
\fmf{plain,tension=1,left=0.25}{v4,vc5}
\fmf{plain,tension=1,left=0.25}{v9,vc7}
\fmf{plain,left=0.25}{w2,vc6}
\fmf{plain,tension=0.25,left=0.25}{v8,vc8}
\fmf{plain,tension=0.25,right=0.25}{v10,vc8}
\fmf{plain,left=0.25}{v7,vc4}
\fmf{phantom,left=0.25}{w2,wc2}
\fmf{plain,tension=0.5}{vc1,vc2}
\fmf{phantom,tension=0.5}{wc1,wc2}
\fmf{plain,tension=0.5}{vc2,vc3}
\fmf{phantom,tension=0.5}{wc2,vc5}
\fmf{plain,tension=0.5}{vc3,vc4}
\fmf{phantom,tension=0.5}{vc4,wc1}
\fmf{plain,tension=0.5}{vc5,vc6}
\fmf{plain,tension=0.5}{vc6,vc7}
\fmf{plain,tension=0.5}{vc7,vc8}
\fmf{plain}{v1,v5}
\fmf{plain,tension=0.5,right=0,width=1mm}{v5,v10}
\fmffreeze
\fmf{dots,tension=0.5}{vc4,vc5}
\fmfposition
\fmfipath{p[]}
\fmfipair{wz[]}
\fmfiset{p1}{vpath(__v6,__vc2)}
\fmfiset{p2}{vpath(__v1,__v5)}
\fmfiset{p3}{vpath(__v10,__vc8)}
\svertex{wz1}{p1}
\vvertex{wz2}{wz1}{p2}
\svertex{wz3}{p2}
\svertex{wz4}{p3}
\fmfi{wiggly}{wz1..wz2}
\wigglywrap{wz3}{v5}{v10}{wz4}
\end{fmfchar*}}
}
\qquad$\cdots$\qquad
\subfigure[$W_{L,L-1}$]{
\fmfframe(3,1)(1,4){%
\begin{fmfchar*}(25,20)
\fmftop{v1}
\fmfbottom{v5}
\fmfforce{(0w,h)}{v1}
\fmfforce{(0w,0)}{v5}
\fmffixed{(0.2w,0)}{v1,v2}
\fmffixed{(0.2w,0)}{v2,v3}
\fmffixed{(0.2w,0)}{v3,w1}
\fmffixed{(0.2w,0)}{w1,v4}
\fmffixed{(0.2w,0)}{v4,v9}
\fmffixed{(0.2w,0)}{v5,v6}
\fmffixed{(0.2w,0)}{v6,v7}
\fmffixed{(0.2w,0)}{v7,w2}
\fmffixed{(0.2w,0)}{w2,v8}
\fmffixed{(0.2w,0)}{v8,v10}
\fmffixed{(0,whatever)}{vc7,vc8}
\fmf{plain,tension=0.25,right=0.25}{v4,vc7}
\fmf{plain,tension=0.25,left=0.25}{v9,vc7}
\fmf{plain,tension=0.25,left=0.25}{v8,vc8}
\fmf{plain,tension=0.25,right=0.25}{v10,vc8}
\fmf{plain}{v2,v6}
\fmf{plain}{w2,w1}
\fmf{plain,tension=0.5}{vc7,vc8}
\fmf{plain}{v1,v5}
\fmf{plain,tension=0.5,right=0,width=1mm}{v5,v10}
\fmffreeze
\fmfposition
\fmfipath{p[]}
\fmfipair{wz[]}
\fmfiset{p1}{vpath(__v2,__v6)}
\fmfiset{p2}{vpath(__w1,__w2)}
\fmfiset{p3}{vpath(__v10,__vc8)}
\fmfiset{p4}{vpath(__v1,__v5)}
\fmfiset{p5}{vpath(__vc8,__vc7)}
\svertex{wz1}{p1}
\vvertex{wz2}{wz1}{p2}
\svertex{wz3}{p3}
\vvertex{wz4}{wz3}{p4}
\fmfiequ{wz5}{point length(p4)/3 of p4}
\fmfiequ{wz6}{point 2length(p1)/3 of p1}
\fmfiequ{wz7}{point length(p2)/3 of p2}
\svertex{wz8}{p5}
\fmfi{wiggly}{wz7..wz8}
\fmfi{dots}{wz1..wz2}
\fmfi{wiggly}{wz5..wz6}
\wigglywrap{wz4}{v5}{v10}{wz3}
\end{fmfchar*}}}
}
\caption{Sample diagrams from different classes}
\label{diagrams-vec}
\end{figure}

Next we have all the wrapping diagrams which contain vector propagators. They can be collected in different sets distinguished by their chiral structure and correspondingly the number of vector lines that enter the graph. One representative for each set is shown in Fig.~\ref{diagrams-vec}. We have:\\
\parbox[h]{\textwidth}{
\begin{itemize}
\addtolength{\itemsep}{-15pt}
\item graphs with chiral structure $\dchi{L-1,\dots,2,1}$ and $1$ vector \\
\item graphs with chiral structure $\dchi{L-2,\dots,2,1} ~+~1$ single Z line,  and 2 vectors \\
\item[]\hspace{0.43\textwidth}$\vdots$\\
\item graphs with chiral structure $\dchi{1}~+~(L-2)$ single Z lines, and $L-1$ vectors.
\end{itemize}
}

Since vectors can attach on the single Z lines via the $V_1$ and $V_2$ vertices in (\ref{vertices}), the number of diagrams that one produces in this way is very, very large. Now we are going to prove that remarkable cancellations do occur. In fact we need only consider graphs with only  $V_2$ vertices on the single Z lines since all the diagrams that contain any $V_1$  vertex on a single Z line do actually cancel out completely.

\begin{figure}[b]
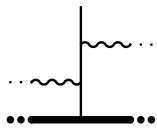

\unitlength=0.75mm
\settoheight{\eqoff}{$\times$}%
\setlength{\eqoff}{0.5\eqoff}%
\addtolength{\eqoff}{-12.5\unitlength}%
\settoheight{\eqofftwo}{$\times$}%
\setlength{\eqofftwo}{0.5\eqofftwo}%
\addtolength{\eqofftwo}{-7.5\unitlength}%
\centering
\raisebox{\eqoff}{%
\fmfframe(3,1)(1,4){%
\begin{fmfchar*}(25,20)
\fmftop{v1}
\fmfbottom{v6}
\fmfforce{(0w,h)}{v1}
\fmfforce{(0w,0)}{v6}
\fmffixed{(0.15w,0)}{v1,v2}
\fmffixed{(0.35w,0)}{v2,v3}
\fmffixed{(0.35w,0)}{v3,v4}
\fmffixed{(0.15w,0)}{v4,v5}
\fmffixed{(0.15w,0)}{v6,v7}
\fmffixed{(0.35w,0)}{v7,v8}
\fmffixed{(0.35w,0)}{v8,v9}
\fmffixed{(0.15w,0)}{v9,v10}
\fmf{plain}{v3,v8}
\fmf{phantom}{v2,v7}
\fmf{phantom}{v1,v6}
\fmf{phantom}{v4,v9}
\fmf{phantom}{v5,v10}
\fmf{plain,tension=0.5,right=0,width=1mm}{v7,v9}
\fmf{dots,tension=0.5,right=0,width=1mm}{v6,v7}
\fmf{dots,tension=0.5,right=0,width=1mm}{v9,v10}
\fmffreeze
\fmfposition
\fmfipath{p[]}
\fmfipair{w[]}
\fmfiset{p1}{vpath(__v3,__v8)}
\fmfiset{p2}{vpath(__v2,__v7)}
\fmfiset{p3}{vpath(__v4,__v9)}
\fmfiset{p4}{vpath(__v1,__v6)}
\fmfiset{p5}{vpath(__v5,__v10)}
\fmfiequ{w1}{point length(p1)/3 of p1}
\fmfiequ{w2}{point 2length(p1)/3 of p1}
\vvertex{w3}{w2}{p2}
\vvertex{w4}{w1}{p3}
\vvertex{w5}{w2}{p4}
\vvertex{w6}{w1}{p5}
\fmfi{wiggly}{w3..w2}
\fmfi{wiggly}{w1..w4}
\fmfi{dots}{w5..w3}
\fmfi{dots}{w4..w6}
\fmfposition
\end{fmfchar*}}
}
\caption{Z line with two $V_1$ vertices}
\label{struct-2vec}
\end{figure}

\begin{figure}[h]
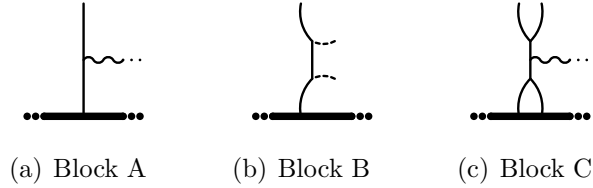

\unitlength=0.75mm
\settoheight{\eqoff}{$\times$}%
\setlength{\eqoff}{0.5\eqoff}%
\addtolength{\eqoff}{-12.5\unitlength}%
\settoheight{\eqofftwo}{$\times$}%
\setlength{\eqofftwo}{0.5\eqofftwo}%
\addtolength{\eqofftwo}{-7.5\unitlength}%
\centering
\raisebox{\eqoff}{%
\subfigure[Block A]{
\label{block-A}
\fmfframe(3,1)(1,4){%
\begin{fmfchar*}(20,20)
\fmftop{v1}
\fmfbottom{v6}
\fmfforce{(0w,h)}{v1}
\fmfforce{(0w,0)}{v6}
\fmffixed{(0.15w,0)}{v1,v2}
\fmffixed{(0.35w,0)}{v2,v3}
\fmffixed{(0.35w,0)}{v3,v4}
\fmffixed{(0.15w,0)}{v4,v5}
\fmffixed{(0.15w,0)}{v6,v7}
\fmffixed{(0.35w,0)}{v7,v8}
\fmffixed{(0.35w,0)}{v8,v9}
\fmffixed{(0.15w,0)}{v9,v10}
\fmf{plain}{v3,v8}
\fmf{phantom}{v2,v7}
\fmf{phantom}{v1,v6}
\fmf{phantom}{v4,v9}
\fmf{phantom}{v5,v10}
\fmf{plain,tension=0.5,right=0,width=1mm}{v7,v9}
\fmf{dots,tension=0.5,right=0,width=1mm}{v6,v7}
\fmf{dots,tension=0.5,right=0,width=1mm}{v9,v10}
\fmffreeze
\fmfposition
\fmfipath{p[]}
\fmfipair{w[]}
\fmfiset{p1}{vpath(__v3,__v8)}
\fmfiset{p2}{vpath(__v2,__v7)}
\fmfiset{p3}{vpath(__v4,__v9)}
\fmfiset{p4}{vpath(__v1,__v6)}
\fmfiset{p5}{vpath(__v5,__v10)}
\fmfiequ{w1}{point length(p1)/2 of p1}
\fmfiequ{w2}{point 2length(p1)/3 of p1}
\vvertex{w4}{w1}{p3}
\vvertex{w6}{w1}{p5}
\fmfi{wiggly}{w1..w4}
\fmfi{dots}{w4..w6}
\fmfposition
\end{fmfchar*}}
}
\qquad
\subfigure[Block B]{
\label{block-B}
\fmfframe(3,1)(1,4){%
\begin{fmfchar*}(20,20)
\fmftop{v1}
\fmfbottom{v7}
\fmfforce{(0w,h)}{v1}
\fmfforce{(0w,0)}{v7}
\fmffixed{(0.15w,0)}{v1,v2}
\fmffixed{(0.25w,0)}{v2,v3}
\fmffixed{(0.2w,0)}{v3,v4}
\fmffixed{(0.25w,0)}{v4,v5}
\fmffixed{(0.15w,0)}{v5,v6}
\fmffixed{(0.15w,0)}{v7,v8}
\fmffixed{(0.25w,0)}{v8,v9}
\fmffixed{(0.2w,0)}{v9,v10}
\fmffixed{(0.25w,0)}{v10,v11}
\fmffixed{(0.15w,0)}{v11,v12}
\fmf{phantom}{v1,v7}
\fmf{phantom}{v2,v8}
\fmf{phantom}{v5,v11}
\fmf{phantom}{v6,v12}
\fmf{plain,tension=0.5,right=0,width=1mm}{v8,v11}
\fmf{dots,tension=0.5,right=0,width=1mm}{v7,v8}
\fmf{dots,tension=0.5,right=0,width=1mm}{v11,v12}
\fmf{plain,tension=0.25,right=0.25}{v3,vc1}
\fmfset{dash_len}{1.5mm}
\fmf{phantom,tension=0.25,left=0.25}{v4,vc1}
\fmf{plain,tension=0.25,left=0.25}{v9,vc2}
\fmf{phantom,tension=0.25,right=0.25}{v10,vc2}
\fmf{plain,tension=0.5}{vc1,vc2}
\fmffreeze
\fmffixed{(0.2w,0)}{vc1,vc3}
\fmffixed{(0.2w,0)}{vc2,vc4}
\fmf{dashes,tension=0.25,left=0.25}{vc3,vc1}
\fmf{dashes,tension=0.25,right=0.25}{vc4,vc2}
\fmfposition
\fmfipath{p[]}
\fmfipair{w[]}
\fmfiset{p1}{vpath(__vc1,__vc2)}
\fmfiset{p2}{vpath(__v5,__v11)}
\fmfiset{p3}{vpath(__v6,__v12)}
\svertex{w1}{p1}
\vvertex{w2}{w1}{p2}
\vvertex{w3}{w1}{p3}
\fmfposition
\end{fmfchar*}}}
\qquad
\subfigure[Block C]{
\label{block-C}
\fmfframe(3,1)(1,4){%
\begin{fmfchar*}(20,20)
\fmftop{v1}
\fmfbottom{v7}
\fmfforce{(0w,h)}{v1}
\fmfforce{(0w,0)}{v7}
\fmffixed{(0.15w,0)}{v1,v2}
\fmffixed{(0.25w,0)}{v2,v3}
\fmffixed{(0.2w,0)}{v3,v4}
\fmffixed{(0.25w,0)}{v4,v5}
\fmffixed{(0.15w,0)}{v5,v6}
\fmffixed{(0.15w,0)}{v7,v8}
\fmffixed{(0.25w,0)}{v8,v9}
\fmffixed{(0.2w,0)}{v9,v10}
\fmffixed{(0.25w,0)}{v10,v11}
\fmffixed{(0.15w,0)}{v11,v12}
\fmf{phantom}{v1,v7}
\fmf{phantom}{v2,v8}
\fmf{phantom}{v5,v11}
\fmf{phantom}{v6,v12}
\fmf{plain,tension=0.5,right=0,width=1mm}{v8,v11}
\fmf{dots,tension=0.5,right=0,width=1mm}{v7,v8}
\fmf{dots,tension=0.5,right=0,width=1mm}{v11,v12}
\fmfset{dash_len}{1.5mm}
\fmf{plain,tension=0.25,right=0.25}{v3,vc1}
\fmf{plain,tension=0.25,left=0.25}{v4,vc1}
\fmf{plain,tension=0.25,left=0.25}{v9,vc2}
\fmf{plain,tension=0.25,right=0.25}{v10,vc2}
\fmf{plain,tension=0.5}{vc1,vc2}
\fmffreeze
\fmfposition
\fmfipath{p[]}
\fmfipair{w[]}
\fmfiset{p1}{vpath(__vc1,__vc2)}
\fmfiset{p2}{vpath(__v5,__v11)}
\fmfiset{p3}{vpath(__v6,__v12)}
\svertex{w1}{p1}
\vvertex{w2}{w1}{p2}
\vvertex{w3}{w1}{p3}
\fmfi{wiggly}{w1..w2}
\fmfi{dots}{w2..w3}
\fmfposition
\end{fmfchar*}}
}
}
\caption{Building blocks for diagrams with vector interactions}
\end{figure}

The proof can be organized as follows:\\
Let us consider a generic diagram where a single Z line has two vectors attaching to it via two $V_1$ vertices as shown in Fig.~\ref{struct-2vec}. This structure appears  in any of the  different sets of diagrams $W_{L,2}, \dots,W_{L,L-1}$ depicted in Fig.~\ref{diagrams-vec}. 
There are only three distinct possibilities to attach the vector on the right hand side of Fig.~\ref{struct-2vec} to the rest of the graph, corresponding to the structures shown in Figs.~\ref{block-A}, \ref{block-B} and \ref{block-C}. 

\begin{figure}[h]
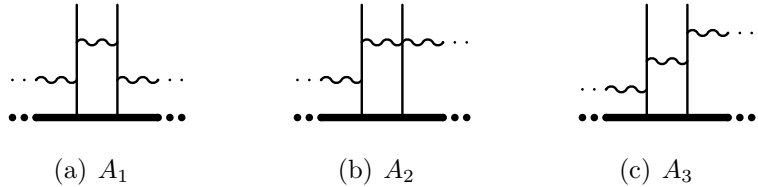

\unitlength=0.75mm
\settoheight{\eqoff}{$\times$}%
\setlength{\eqoff}{0.5\eqoff}%
\addtolength{\eqoff}{-12.5\unitlength}%
\settoheight{\eqofftwo}{$\times$}%
\setlength{\eqofftwo}{0.5\eqofftwo}%
\addtolength{\eqofftwo}{-7.5\unitlength}%
\centering
\raisebox{\eqoff}{%
\subfigure[$A_1$]{
\fmfframe(3,1)(1,4){%
\begin{fmfchar*}(30,20)
\fmftop{v1}
\fmfbottom{v7}
\fmfforce{(0w,h)}{v1}
\fmfforce{(0w,0)}{v7}
\fmffixed{(0.14w,0)}{v1,v2}
\fmffixed{(0.24w,0)}{v2,v3}
\fmffixed{(0.24w,0)}{v3,v4}
\fmffixed{(0.24w,0)}{v4,v5}
\fmffixed{(0.14w,0)}{v5,v6}
\fmffixed{(0.14w,0)}{v7,v8}
\fmffixed{(0.24w,0)}{v8,v9}
\fmffixed{(0.24w,0)}{v9,v10}
\fmffixed{(0.24w,0)}{v10,v11}
\fmffixed{(0.14w,0)}{v11,v12}
\fmf{phantom}{v1,v7}
\fmf{phantom}{v2,v8}
\fmf{plain}{v3,v9}
\fmf{plain}{v4,v10}
\fmf{phantom}{v5,v11}
\fmf{phantom}{v6,v12}
\fmf{plain,tension=0.5,right=0,width=1mm}{v8,v11}
\fmf{dots,tension=0.5,right=0,width=1mm}{v7,v8}
\fmf{dots,tension=0.5,right=0,width=1mm}{v11,v12}
\fmffreeze
\fmfposition
\fmfipath{p[]}
\fmfipair{w[]}
\fmfiset{p1}{vpath(__v1,__v7)}
\fmfiset{p2}{vpath(__v2,__v8)}
\fmfiset{p3}{vpath(__v3,__v9)}
\fmfiset{p4}{vpath(__v4,__v10)}
\fmfiset{p5}{vpath(__v5,__v11)}
\fmfiset{p6}{vpath(__v6,__v12)}
\fmfiequ{w1}{point length(p3)/3 of p3}
\fmfiequ{w2}{point 2length(p3)/3 of p3}
\vvertex{w3}{w2}{p1}
\vvertex{w4}{w2}{p2}
\vvertex{w5}{w1}{p4}
\vvertex{w6}{w2}{p5}
\vvertex{w7}{w2}{p6}
\vvertex{w8}{w2}{p4}
\fmfi{wiggly}{w1..w5}
\fmfi{wiggly}{w4..w2}
\fmfi{wiggly}{w8..w6}
\fmfi{dots}{w3..w4}
\fmfi{dots}{w6..w7}
\fmfposition
\end{fmfchar*}}
}
\qquad
\subfigure[$A_2$]{
\fmfframe(3,1)(1,4){%
\begin{fmfchar*}(30,20)
\fmftop{v1}
\fmfbottom{v7}
\fmfforce{(0w,h)}{v1}
\fmfforce{(0w,0)}{v7}
\fmffixed{(0.14w,0)}{v1,v2}
\fmffixed{(0.24w,0)}{v2,v3}
\fmffixed{(0.24w,0)}{v3,v4}
\fmffixed{(0.24w,0)}{v4,v5}
\fmffixed{(0.14w,0)}{v5,v6}
\fmffixed{(0.14w,0)}{v7,v8}
\fmffixed{(0.24w,0)}{v8,v9}
\fmffixed{(0.24w,0)}{v9,v10}
\fmffixed{(0.24w,0)}{v10,v11}
\fmffixed{(0.14w,0)}{v11,v12}
\fmf{phantom}{v1,v7}
\fmf{phantom}{v2,v8}
\fmf{plain}{v3,v9}
\fmf{plain}{v4,v10}
\fmf{phantom}{v5,v11}
\fmf{phantom}{v6,v12}
\fmf{plain,tension=0.5,right=0,width=1mm}{v8,v11}
\fmf{dots,tension=0.5,right=0,width=1mm}{v7,v8}
\fmf{dots,tension=0.5,right=0,width=1mm}{v11,v12}
\fmffreeze
\fmfposition
\fmfipath{p[]}
\fmfipair{w[]}
\fmfiset{p1}{vpath(__v1,__v7)}
\fmfiset{p2}{vpath(__v2,__v8)}
\fmfiset{p3}{vpath(__v3,__v9)}
\fmfiset{p4}{vpath(__v4,__v10)}
\fmfiset{p5}{vpath(__v5,__v11)}
\fmfiset{p6}{vpath(__v6,__v12)}
\fmfiequ{w1}{point length(p3)/3 of p3}
\fmfiequ{w2}{point 2length(p3)/3 of p3}
\vvertex{w3}{w2}{p1}
\vvertex{w4}{w2}{p2}
\vvertex{w5}{w1}{p4}
\vvertex{w6}{w1}{p5}
\vvertex{w7}{w1}{p6}
\vvertex{w8}{w1}{p4}
\fmfi{wiggly}{w1..w5}
\fmfi{wiggly}{w4..w2}
\fmfi{wiggly}{w8..w6}
\fmfi{dots}{w3..w4}
\fmfi{dots}{w6..w7}
\fmfposition
\end{fmfchar*}}
}
\qquad
\subfigure[$A_3$]{
\fmfframe(3,1)(1,4){%
\begin{fmfchar*}(30,20)
\fmftop{v1}
\fmfbottom{v7}
\fmfforce{(0w,h)}{v1}
\fmfforce{(0w,0)}{v7}
\fmffixed{(0.14w,0)}{v1,v2}
\fmffixed{(0.24w,0)}{v2,v3}
\fmffixed{(0.24w,0)}{v3,v4}
\fmffixed{(0.24w,0)}{v4,v5}
\fmffixed{(0.14w,0)}{v5,v6}
\fmffixed{(0.14w,0)}{v7,v8}
\fmffixed{(0.24w,0)}{v8,v9}
\fmffixed{(0.24w,0)}{v9,v10}
\fmffixed{(0.24w,0)}{v10,v11}
\fmffixed{(0.14w,0)}{v11,v12}
\fmf{phantom}{v1,v7}
\fmf{phantom}{v2,v8}
\fmf{plain}{v3,v9}
\fmf{plain}{v4,v10}
\fmf{phantom}{v5,v11}
\fmf{phantom}{v6,v12}
\fmf{plain,tension=0.5,right=0,width=1mm}{v8,v11}
\fmf{dots,tension=0.5,right=0,width=1mm}{v7,v8}
\fmf{dots,tension=0.5,right=0,width=1mm}{v11,v12}
\fmffreeze
\fmfposition
\fmfipath{p[]}
\fmfipair{w[]}
\fmfiset{p1}{vpath(__v1,__v7)}
\fmfiset{p2}{vpath(__v2,__v8)}
\fmfiset{p3}{vpath(__v3,__v9)}
\fmfiset{p4}{vpath(__v4,__v10)}
\fmfiset{p5}{vpath(__v5,__v11)}
\fmfiset{p6}{vpath(__v6,__v12)}
\fmfiequ{w1}{point length(p3)/4 of p3}
\fmfiequ{w2}{point length(p3)/2 of p3}
\fmfiequ{w3}{point 3length(p3)/4 of p3}
\vvertex{w4}{w3}{p1}
\vvertex{w5}{w3}{p2}
\vvertex{w6}{w2}{p4}
\vvertex{w7}{w1}{p4}
\vvertex{w8}{w1}{p5}
\vvertex{w9}{w1}{p6}
\fmfi{wiggly}{w5..w3}
\fmfi{wiggly}{w2..w6}
\fmfi{wiggly}{w7..w8}
\fmfi{dots}{w4..w5}
\fmfi{dots}{w8..w9}
\fmfposition
\end{fmfchar*}}}
}

\caption{Diagrams of class A}
\label{diagrams-A}
\end{figure}

\begin{figure}[h]
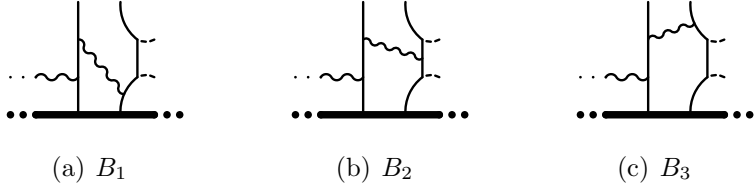

\unitlength=0.75mm
\settoheight{\eqoff}{$\times$}%
\setlength{\eqoff}{0.5\eqoff}%
\addtolength{\eqoff}{-12.5\unitlength}%
\settoheight{\eqofftwo}{$\times$}%
\setlength{\eqofftwo}{0.5\eqofftwo}%
\addtolength{\eqofftwo}{-7.5\unitlength}%
\centering
\raisebox{\eqoff}{%
\subfigure[$B_1$]{
\fmfframe(3,1)(1,4){%
\begin{fmfchar*}(30,20)
\fmftop{v1}
\fmfbottom{v7}
\fmfforce{(0w,h)}{v1}
\fmfforce{(0w,0)}{v7}
\fmffixed{(0.15w,0)}{v1,v2}
\fmffixed{(0.25w,0)}{v2,v3}
\fmffixed{(0.25w,0)}{v3,v4}
\fmffixed{(0.2w,0)}{v4,v5}
\fmffixed{(0.15w,0)}{v5,v6}
\fmffixed{(0.15w,0)}{v7,v8}
\fmffixed{(0.25w,0)}{v8,v9}
\fmffixed{(0.25w,0)}{v9,v10}
\fmffixed{(0.2w,0)}{v10,v11}
\fmffixed{(0.15w,0)}{v11,v12}
\fmf{phantom}{v1,v7}
\fmf{phantom}{v2,v8}
\fmf{plain}{v3,v9}
\fmfset{dash_len}{1.5mm}
\fmf{plain,tension=0.25,right=0.25}{v4,vc1}
\fmf{phantom,tension=0.25,left=0.25}{v5,vc1}
\fmf{plain,tension=0.25,left=0.25}{v10,vc2}
\fmf{phantom,tension=0.25,right=0.25}{v11,vc2}
\fmf{plain,tension=0.5}{vc1,vc2}
\fmf{plain,tension=0.5,right=0,width=1mm}{v8,v11}
\fmf{dots,tension=0.5,right=0,width=1mm}{v7,v8}
\fmf{dots,tension=0.5,right=0,width=1mm}{v11,v12}
\fmffreeze
\fmfposition
\fmffixed{(0.1w,0)}{vc1,vc3}
\fmffixed{(0.1w,0)}{vc2,vc4}
\fmf{dashes,tension=0.25,left=0.25}{vc3,vc1}
\fmf{dashes,tension=0.25,right=0.25}{vc4,vc2}
\fmfipath{p[]}
\fmfipair{w[]}
\fmfiset{p1}{vpath(__v1,__v7)}
\fmfiset{p2}{vpath(__v2,__v8)}
\fmfiset{p3}{vpath(__v3,__v9)}
\fmfiset{p4}{vpath(__vc2,__v10)}
\fmfiequ{w1}{point length(p3)/3 of p3}
\fmfiequ{w2}{point 2length(p3)/3 of p3}
\vvertex{w3}{w2}{p1}
\vvertex{w4}{w2}{p2}
\svertex{w5}{p4}
\fmfi{wiggly}{w4..w2}
\fmfi{wiggly}{w1..w5}
\fmfi{dots}{w3..w4}
\fmfposition
\end{fmfchar*}}
}
\qquad
\subfigure[$B_2$]{
\fmfframe(3,1)(1,4){%
\begin{fmfchar*}(30,20)
\fmftop{v1}
\fmfbottom{v7}
\fmfforce{(0w,h)}{v1}
\fmfforce{(0w,0)}{v7}
\fmffixed{(0.15w,0)}{v1,v2}
\fmffixed{(0.25w,0)}{v2,v3}
\fmffixed{(0.25w,0)}{v3,v4}
\fmffixed{(0.2w,0)}{v4,v5}
\fmffixed{(0.15w,0)}{v5,v6}
\fmffixed{(0.15w,0)}{v7,v8}
\fmffixed{(0.25w,0)}{v8,v9}
\fmffixed{(0.25w,0)}{v9,v10}
\fmffixed{(0.2w,0)}{v10,v11}
\fmffixed{(0.15w,0)}{v11,v12}
\fmf{phantom}{v1,v7}
\fmf{phantom}{v2,v8}
\fmf{plain}{v3,v9}
\fmfset{dash_len}{1.5mm}
\fmf{plain,tension=0.25,right=0.25}{v4,vc1}
\fmf{phantom,tension=0.25,left=0.25}{v5,vc1}
\fmf{plain,tension=0.25,left=0.25}{v10,vc2}
\fmf{phantom,tension=0.25,right=0.25}{v11,vc2}
\fmf{plain,tension=0.5}{vc1,vc2}
\fmf{plain,tension=0.5,right=0,width=1mm}{v8,v11}
\fmf{dots,tension=0.5,right=0,width=1mm}{v7,v8}
\fmf{dots,tension=0.5,right=0,width=1mm}{v11,v12}
\fmffreeze
\fmfposition
\fmffixed{(0.1w,0)}{vc1,vc3}
\fmffixed{(0.1w,0)}{vc2,vc4}
\fmf{dashes,tension=0.25,left=0.25}{vc3,vc1}
\fmf{dashes,tension=0.25,right=0.25}{vc4,vc2}
\fmfipath{p[]}
\fmfipair{w[]}
\fmfiset{p1}{vpath(__v1,__v7)}
\fmfiset{p2}{vpath(__v2,__v8)}
\fmfiset{p3}{vpath(__v3,__v9)}
\fmfiset{p4}{vpath(__vc2,__vc1)}
\fmfiequ{w1}{point length(p3)/3 of p3}
\fmfiequ{w2}{point 2length(p3)/3 of p3}
\vvertex{w3}{w2}{p1}
\vvertex{w4}{w2}{p2}
\svertex{w5}{p4}
\fmfi{wiggly}{w4..w2}
\fmfi{wiggly}{w1..w5}
\fmfi{dots}{w3..w4}
\fmfposition
\end{fmfchar*}}
}
\qquad
\subfigure[$B_3$]{
\fmfframe(3,1)(1,4){%
\begin{fmfchar*}(30,20)
\fmftop{v1}
\fmfbottom{v7}
\fmfforce{(0w,h)}{v1}
\fmfforce{(0w,0)}{v7}
\fmffixed{(0.15w,0)}{v1,v2}
\fmffixed{(0.25w,0)}{v2,v3}
\fmffixed{(0.25w,0)}{v3,v4}
\fmffixed{(0.2w,0)}{v4,v5}
\fmffixed{(0.15w,0)}{v5,v6}
\fmffixed{(0.15w,0)}{v7,v8}
\fmffixed{(0.25w,0)}{v8,v9}
\fmffixed{(0.25w,0)}{v9,v10}
\fmffixed{(0.2w,0)}{v10,v11}
\fmffixed{(0.15w,0)}{v11,v12}
\fmf{phantom}{v1,v7}
\fmf{phantom}{v2,v8}
\fmf{plain}{v3,v9}
\fmfset{dash_len}{1.5mm}
\fmf{plain,tension=0.25,right=0.25}{v4,vc1}
\fmf{phantom,tension=0.25,left=0.25}{v5,vc1}
\fmf{plain,tension=0.25,left=0.25}{v10,vc2}
\fmf{phantom,tension=0.25,right=0.25}{v11,vc2}
\fmf{plain,tension=0.5}{vc1,vc2}
\fmf{plain,tension=0.5,right=0,width=1mm}{v8,v11}
\fmf{dots,tension=0.5,right=0,width=1mm}{v7,v8}
\fmf{dots,tension=0.5,right=0,width=1mm}{v11,v12}
\fmffreeze
\fmfposition
\fmffixed{(0.1w,0)}{vc1,vc3}
\fmffixed{(0.1w,0)}{vc2,vc4}
\fmf{dashes,tension=0.25,left=0.25}{vc3,vc1}
\fmf{dashes,tension=0.25,right=0.25}{vc4,vc2}
\fmfipath{p[]}
\fmfipair{w[]}
\fmfiset{p1}{vpath(__v1,__v7)}
\fmfiset{p2}{vpath(__v2,__v8)}
\fmfiset{p3}{vpath(__v3,__v9)}
\fmfiset{p4}{vpath(__v4,__vc1)}
\fmfiequ{w1}{point length(p3)/3 of p3}
\fmfiequ{w2}{point 2length(p3)/3 of p3}
\vvertex{w3}{w2}{p1}
\vvertex{w4}{w2}{p2}
\svertex{w5}{p4}
\fmfi{wiggly}{w4..w2}
\fmfi{wiggly}{w1..w5}
\fmfi{dots}{w3..w4}
\fmfposition
\end{fmfchar*}}}
}
\caption{Diagrams of class B}
\label{diagrams-B}
\end{figure}

\begin{figure}[h]
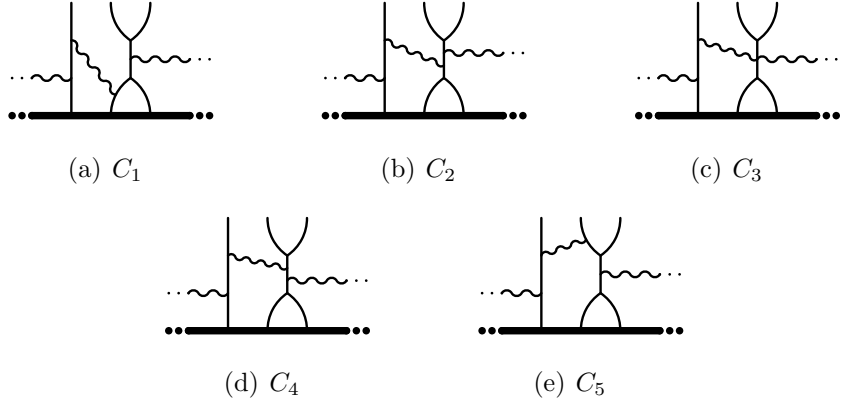

\unitlength=0.75mm
\settoheight{\eqoff}{$\times$}%
\setlength{\eqoff}{0.5\eqoff}%
\addtolength{\eqoff}{-12.5\unitlength}%
\settoheight{\eqofftwo}{$\times$}%
\setlength{\eqofftwo}{0.5\eqofftwo}%
\addtolength{\eqofftwo}{-7.5\unitlength}%
\centering
\raisebox{\eqoff}{%
\subfigure[$C_1$]{
\fmfframe(3,1)(1,4){%
\begin{fmfchar*}(35,20)
\fmftop{v1}
\fmfbottom{v8}
\fmfforce{(0w,h)}{v1}
\fmfforce{(0w,0)}{v8}
\fmffixed{(0.1w,0)}{v1,v2}
\fmffixed{(0.2w,0)}{v2,v3}
\fmffixed{(0.2w,0)}{v3,v4}
\fmffixed{(0.2w,0)}{v4,v5}
\fmffixed{(0.2w,0)}{v5,v6}
\fmffixed{(0.1w,0)}{v6,v7}
\fmffixed{(0.1w,0)}{v8,v9}
\fmffixed{(0.2w,0)}{v9,v10}
\fmffixed{(0.2w,0)}{v10,v11}
\fmffixed{(0.2w,0)}{v11,v12}
\fmffixed{(0.2w,0)}{v12,v13}
\fmffixed{(0.1w,0)}{v13,v14}
\fmf{phantom}{v1,v8}
\fmf{phantom}{v2,v9}
\fmf{plain}{v3,v10}
\fmf{phantom}{v6,v13}
\fmf{phantom}{v7,v14}
\fmf{plain,tension=0.25,right=0.25}{v4,vc1}
\fmf{plain,tension=0.25,left=0.25}{v5,vc1}
\fmf{plain,tension=0.25,left=0.25}{v11,vc2}
\fmf{plain,tension=0.25,right=0.25}{v12,vc2}
\fmf{plain,tension=0.5}{vc1,vc2}
\fmf{plain,tension=0.5,right=0,width=1mm}{v9,v13}
\fmf{dots,tension=0.5,right=0,width=1mm}{v8,v9}
\fmf{dots,tension=0.5,right=0,width=1mm}{v13,v14}
\fmffreeze
\fmfposition
\fmfipath{p[]}
\fmfipair{w[]}
\fmfiset{p1}{vpath(__v1,__v8)}
\fmfiset{p2}{vpath(__v2,__v9)}
\fmfiset{p3}{vpath(__v3,__v10)}
\fmfiset{p4}{vpath(__v11,__vc2)}
\fmfiset{p5}{vpath(__v6,__v13)}
\fmfiset{p6}{vpath(__v7,__v14)}
\fmfiset{p7}{vpath(__vc1,__vc2)}
\fmfiequ{w1}{point length(p3)/3 of p3}
\fmfiequ{w2}{point 2length(p3)/3 of p3}
\vvertex{w3}{w2}{p1}
\vvertex{w4}{w2}{p2}
\svertex{w5}{p4}
\svertex{w6}{p7}
\vvertex{w7}{w6}{p5}
\vvertex{w8}{w6}{p6}
\fmfi{wiggly}{w4..w2}
\fmfi{wiggly}{w1..w5}
\fmfi{dots}{w3..w4}
\fmfi{wiggly}{w6..w7}
\fmfi{dots}{w7..w8}
\fmfposition
\end{fmfchar*}}
}
\qquad
\subfigure[$C_2$]{
\fmfframe(3,1)(1,4){%
\begin{fmfchar*}(35,20)
\fmftop{v1}
\fmfbottom{v8}
\fmfforce{(0w,h)}{v1}
\fmfforce{(0w,0)}{v8}
\fmffixed{(0.1w,0)}{v1,v2}
\fmffixed{(0.2w,0)}{v2,v3}
\fmffixed{(0.2w,0)}{v3,v4}
\fmffixed{(0.2w,0)}{v4,v5}
\fmffixed{(0.2w,0)}{v5,v6}
\fmffixed{(0.1w,0)}{v6,v7}
\fmffixed{(0.1w,0)}{v8,v9}
\fmffixed{(0.2w,0)}{v9,v10}
\fmffixed{(0.2w,0)}{v10,v11}
\fmffixed{(0.2w,0)}{v11,v12}
\fmffixed{(0.2w,0)}{v12,v13}
\fmffixed{(0.1w,0)}{v13,v14}
\fmf{phantom}{v1,v8}
\fmf{phantom}{v2,v9}
\fmf{plain}{v3,v10}
\fmf{phantom}{v6,v13}
\fmf{phantom}{v7,v14}
\fmf{plain,tension=0.25,right=0.25}{v4,vc1}
\fmf{plain,tension=0.25,left=0.25}{v5,vc1}
\fmf{plain,tension=0.25,left=0.25}{v11,vc2}
\fmf{plain,tension=0.25,right=0.25}{v12,vc2}
\fmf{plain,tension=0.5}{vc1,vc2}
\fmf{plain,tension=0.5,right=0,width=1mm}{v9,v13}
\fmf{dots,tension=0.5,right=0,width=1mm}{v8,v9}
\fmf{dots,tension=0.5,right=0,width=1mm}{v13,v14}
\fmffreeze
\fmfposition
\fmfipath{p[]}
\fmfipair{w[]}
\fmfiset{p1}{vpath(__v1,__v8)}
\fmfiset{p2}{vpath(__v2,__v9)}
\fmfiset{p3}{vpath(__v3,__v10)}
\fmfiset{p4}{vpath(__v11,__vc2)}
\fmfiset{p5}{vpath(__v6,__v13)}
\fmfiset{p6}{vpath(__v7,__v14)}
\fmfiset{p7}{vpath(__vc1,__vc2)}
\fmfiequ{w1}{point length(p3)/3 of p3}
\fmfiequ{w2}{point 2length(p3)/3 of p3}
\vvertex{w3}{w2}{p1}
\vvertex{w4}{w2}{p2}
\fmfiequ{w5}{point 2length(p3)/3 of p7}
\fmfiequ{w6}{point length(p3)/3 of p7}
\vvertex{w7}{w6}{p5}
\vvertex{w8}{w6}{p6}
\fmfi{wiggly}{w4..w2}
\fmfi{wiggly}{w1..w5}
\fmfi{dots}{w3..w4}
\fmfi{wiggly}{w6..w7}
\fmfi{dots}{w7..w8}
\fmfposition
\end{fmfchar*}}
}
\qquad
\subfigure[$C_3$]{
\fmfframe(3,1)(1,4){%
\begin{fmfchar*}(35,20)
\fmftop{v1}
\fmfbottom{v8}
\fmfforce{(0w,h)}{v1}
\fmfforce{(0w,0)}{v8}
\fmffixed{(0.1w,0)}{v1,v2}
\fmffixed{(0.2w,0)}{v2,v3}
\fmffixed{(0.2w,0)}{v3,v4}
\fmffixed{(0.2w,0)}{v4,v5}
\fmffixed{(0.2w,0)}{v5,v6}
\fmffixed{(0.1w,0)}{v6,v7}
\fmffixed{(0.1w,0)}{v8,v9}
\fmffixed{(0.2w,0)}{v9,v10}
\fmffixed{(0.2w,0)}{v10,v11}
\fmffixed{(0.2w,0)}{v11,v12}
\fmffixed{(0.2w,0)}{v12,v13}
\fmffixed{(0.1w,0)}{v13,v14}
\fmf{phantom}{v1,v8}
\fmf{phantom}{v2,v9}
\fmf{plain}{v3,v10}
\fmf{phantom}{v6,v13}
\fmf{phantom}{v7,v14}
\fmf{plain,tension=0.25,right=0.25}{v4,vc1}
\fmf{plain,tension=0.25,left=0.25}{v5,vc1}
\fmf{plain,tension=0.25,left=0.25}{v11,vc2}
\fmf{plain,tension=0.25,right=0.25}{v12,vc2}
\fmf{plain,tension=0.5}{vc1,vc2}
\fmf{plain,tension=0.5,right=0,width=1mm}{v9,v13}
\fmf{dots,tension=0.5,right=0,width=1mm}{v8,v9}
\fmf{dots,tension=0.5,right=0,width=1mm}{v13,v14}
\fmffreeze
\fmfposition
\fmfipath{p[]}
\fmfipair{w[]}
\fmfiset{p1}{vpath(__v1,__v8)}
\fmfiset{p2}{vpath(__v2,__v9)}
\fmfiset{p3}{vpath(__v3,__v10)}
\fmfiset{p4}{vpath(__v11,__vc2)}
\fmfiset{p5}{vpath(__v6,__v13)}
\fmfiset{p6}{vpath(__v7,__v14)}
\fmfiset{p7}{vpath(__vc1,__vc2)}
\fmfiequ{w1}{point length(p3)/3 of p3}
\fmfiequ{w2}{point 2length(p3)/3 of p3}
\vvertex{w3}{w2}{p1}
\vvertex{w4}{w2}{p2}
\fmfiequ{w5}{point length(p3)/2 of p7}
\fmfiequ{w6}{point length(p3)/2 of p7}
\vvertex{w7}{w6}{p5}
\vvertex{w8}{w6}{p6}
\fmfi{wiggly}{w4..w2}
\fmfi{wiggly}{w1..w5}
\fmfi{dots}{w3..w4}
\fmfi{wiggly}{w6..w7}
\fmfi{dots}{w7..w8}
\fmfposition
\end{fmfchar*}}}
}
\\
%
\raisebox{\eqoff}{%
\subfigure[$C_4$]{
\fmfframe(3,1)(1,4){%
\begin{fmfchar*}(35,20)
\fmftop{v1}
\fmfbottom{v8}
\fmfforce{(0w,h)}{v1}
\fmfforce{(0w,0)}{v8}
\fmffixed{(0.1w,0)}{v1,v2}
\fmffixed{(0.2w,0)}{v2,v3}
\fmffixed{(0.2w,0)}{v3,v4}
\fmffixed{(0.2w,0)}{v4,v5}
\fmffixed{(0.2w,0)}{v5,v6}
\fmffixed{(0.1w,0)}{v6,v7}
\fmffixed{(0.1w,0)}{v8,v9}
\fmffixed{(0.2w,0)}{v9,v10}
\fmffixed{(0.2w,0)}{v10,v11}
\fmffixed{(0.2w,0)}{v11,v12}
\fmffixed{(0.2w,0)}{v12,v13}
\fmffixed{(0.1w,0)}{v13,v14}
\fmf{phantom}{v1,v8}
\fmf{phantom}{v2,v9}
\fmf{plain}{v3,v10}
\fmf{phantom}{v6,v13}
\fmf{phantom}{v7,v14}
\fmf{plain,tension=0.25,right=0.25}{v4,vc1}
\fmf{plain,tension=0.25,left=0.25}{v5,vc1}
\fmf{plain,tension=0.25,left=0.25}{v11,vc2}
\fmf{plain,tension=0.25,right=0.25}{v12,vc2}
\fmf{plain,tension=0.5}{vc1,vc2}
\fmf{plain,tension=0.5,right=0,width=1mm}{v9,v13}
\fmf{dots,tension=0.5,right=0,width=1mm}{v8,v9}
\fmf{dots,tension=0.5,right=0,width=1mm}{v13,v14}
\fmffreeze
\fmfposition
\fmfipath{p[]}
\fmfipair{w[]}
\fmfiset{p1}{vpath(__v1,__v8)}
\fmfiset{p2}{vpath(__v2,__v9)}
\fmfiset{p3}{vpath(__v3,__v10)}
\fmfiset{p4}{vpath(__v11,__vc2)}
\fmfiset{p5}{vpath(__v6,__v13)}
\fmfiset{p6}{vpath(__v7,__v14)}
\fmfiset{p7}{vpath(__vc1,__vc2)}
\fmfiequ{w1}{point length(p3)/3 of p3}
\fmfiequ{w2}{point 2length(p3)/3 of p3}
\vvertex{w3}{w2}{p1}
\vvertex{w4}{w2}{p2}
\fmfiequ{w5}{point length(p3)/3 of p7}
\fmfiequ{w6}{point 2length(p3)/3 of p7}
\vvertex{w7}{w6}{p5}
\vvertex{w8}{w6}{p6}
\fmfi{wiggly}{w4..w2}
\fmfi{wiggly}{w1..w5}
\fmfi{dots}{w3..w4}
\fmfi{wiggly}{w6..w7}
\fmfi{dots}{w7..w8}
\fmfposition
\end{fmfchar*}}
}
\qquad
\subfigure[$C_5$]{
\fmfframe(3,1)(1,4){%
\begin{fmfchar*}(35,20)
\fmftop{v1}
\fmfbottom{v8}
\fmfforce{(0w,h)}{v1}
\fmfforce{(0w,0)}{v8}
\fmffixed{(0.1w,0)}{v1,v2}
\fmffixed{(0.2w,0)}{v2,v3}
\fmffixed{(0.2w,0)}{v3,v4}
\fmffixed{(0.2w,0)}{v4,v5}
\fmffixed{(0.2w,0)}{v5,v6}
\fmffixed{(0.1w,0)}{v6,v7}
\fmffixed{(0.1w,0)}{v8,v9}
\fmffixed{(0.2w,0)}{v9,v10}
\fmffixed{(0.2w,0)}{v10,v11}
\fmffixed{(0.2w,0)}{v11,v12}
\fmffixed{(0.2w,0)}{v12,v13}
\fmffixed{(0.1w,0)}{v13,v14}
\fmf{phantom}{v1,v8}
\fmf{phantom}{v2,v9}
\fmf{plain}{v3,v10}
\fmf{phantom}{v6,v13}
\fmf{phantom}{v7,v14}
\fmf{plain,tension=0.25,right=0.25}{v4,vc1}
\fmf{plain,tension=0.25,left=0.25}{v5,vc1}
\fmf{plain,tension=0.25,left=0.25}{v11,vc2}
\fmf{plain,tension=0.25,right=0.25}{v12,vc2}
\fmf{plain,tension=0.5}{vc1,vc2}
\fmf{plain,tension=0.5,right=0,width=1mm}{v9,v13}
\fmf{dots,tension=0.5,right=0,width=1mm}{v8,v9}
\fmf{dots,tension=0.5,right=0,width=1mm}{v13,v14}
\fmffreeze
\fmfposition
\fmfipath{p[]}
\fmfipair{w[]}
\fmfiset{p1}{vpath(__v1,__v8)}
\fmfiset{p2}{vpath(__v2,__v9)}
\fmfiset{p3}{vpath(__v3,__v10)}
\fmfiset{p4}{vpath(__v11,__vc2)}
\fmfiset{p5}{vpath(__v6,__v13)}
\fmfiset{p6}{vpath(__v7,__v14)}
\fmfiset{p7}{vpath(__vc1,__vc2)}
\fmfiset{p8}{vpath(__v4,__vc1)}
\fmfiequ{w1}{point length(p3)/3 of p3}
\fmfiequ{w2}{point 2length(p3)/3 of p3}
\vvertex{w3}{w2}{p1}
\vvertex{w4}{w2}{p2}
\svertex{w5}{p8}
\fmfiequ{w6}{point length(p3)/2 of p7}
\vvertex{w7}{w6}{p5}
\vvertex{w8}{w6}{p6}
\fmfi{wiggly}{w4..w2}
\fmfi{wiggly}{w1..w5}
\fmfi{dots}{w3..w4}
\fmfi{wiggly}{w6..w7}
\fmfi{dots}{w7..w8}
\fmfposition
\end{fmfchar*}}}
}
\caption{Diagrams of class C}
\label{diagrams-C}
\end{figure}

We examine the three situations separately and prove the complete cancellation of the divergent contributions.
 All the possible diagrams coming from each class are shown in Figs.~\ref{diagrams-A}, \ref{diagrams-B} and \ref{diagrams-C}.
\begin{itemize}
\item Class A (Fig.~\ref{diagrams-A}):\\
The diagram $A_1$ is finite.\\ 
Performing part of the D-algebra for the diagram $A_3$ one immediately obtains  the same structure of the diagram $A_2$, with a different sign due to the $\Box=-p^2$ cancelling one propagator. Since the two diagrams have the same color factor, their divergent parts sum up to zero.
\item Class B (Fig.~\ref{diagrams-B}):\\
For the diagrams $B_1$ and $B_2$  a partial D-algebra shows that they  produce the same result 
while the opposite color factors make the total divergent part vanish.\\
The diagram $B_3$ is finite.
\item Class C (Fig.~\ref{diagrams-C}):\\
For diagrams $C_1$ and $C_2$ we have the same situation as for $B_1$ and $B_2$ and the divergent part of the two diagrams cancels out.\\
For the diagrams $C_3$ and $C_4$ the cancellation occurs following the same pattern as for the diagrams $A_2$ and $A_3$.\\
The diagram $C_5$ is finite.
\end{itemize}


\begin{figure}[t]
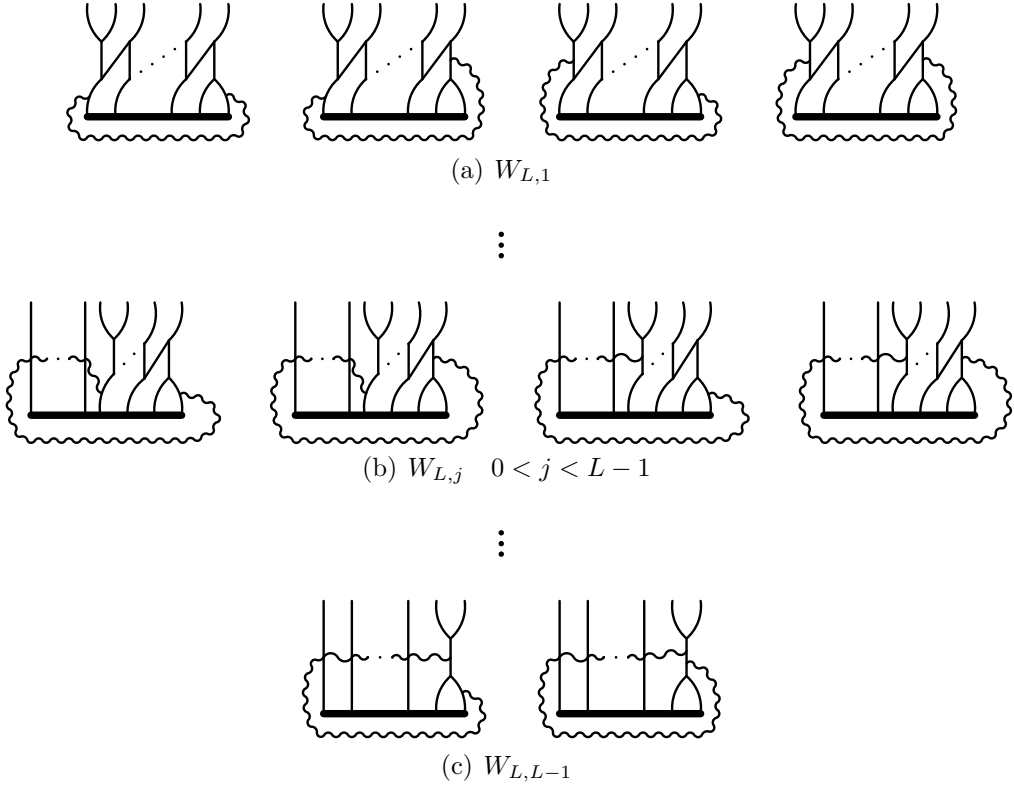

\unitlength=0.75mm
\settoheight{\eqoff}{$\times$}%
\setlength{\eqoff}{0.5\eqoff}%
\addtolength{\eqoff}{-12.5\unitlength}%
\settoheight{\eqofftwo}{$\times$}%
\setlength{\eqofftwo}{0.5\eqofftwo}%
\addtolength{\eqofftwo}{-7.5\unitlength}%
\centering
\raisebox{\eqoff}{%
\subfigure[$W_{L,1}$]
{
\fmfframe(3,1)(1,4){%
\begin{fmfchar*}(25,20)
\fmftop{v1}
\fmfbottom{v5}
\fmfforce{(0w,h)}{v1}
\fmfforce{(0w,0)}{v5}
\fmffixed{(0.2w,0)}{v1,v2}
\fmffixed{(0.2w,0)}{v2,v3}
\fmffixed{(0.2w,0)}{v3,w1}
\fmffixed{(0.2w,0)}{w1,v4}
\fmffixed{(0.2w,0)}{v4,v9}
\fmffixed{(0.2w,0)}{v5,v6}
\fmffixed{(0.2w,0)}{v6,v7}
\fmffixed{(0.2w,0)}{v7,w2}
\fmffixed{(0.2w,0)}{w2,v8}
\fmffixed{(0.2w,0)}{v8,v10}
\fmffixed{(0,whatever)}{vc1,vc2}
\fmffixed{(0,whatever)}{vc3,vc4}
\fmffixed{(0,whatever)}{vc5,vc6}
\fmffixed{(0,whatever)}{vc7,vc8}
\fmf{plain,tension=0.25,right=0.25}{v1,vc1}
\fmf{plain,tension=0.25,left=0.25}{v2,vc1}
\fmf{plain,left=0.25}{v5,vc2}
\fmf{plain,tension=1,left=0.25}{v3,vc3}
\fmf{phantom,tension=1,left=0.25}{w1,wc1}
\fmf{plain,tension=1,left=0.25}{v4,vc5}
\fmf{plain,tension=1,left=0.25}{v9,vc7}
\fmf{plain,left=0.25}{w2,vc6}
\fmf{plain,tension=0.25,left=0.25}{v8,vc8}
\fmf{plain,tension=0.25,right=0.25}{v10,vc8}
\fmf{plain,left=0.25}{v6,vc4}
\fmf{phantom,left=0.25}{v7,wc2}
\fmf{plain,tension=0.5}{vc1,vc2}
\fmf{phantom,tension=0.5}{wc1,wc2}
\fmf{plain,tension=0.5}{vc2,vc3}
\fmf{phantom,tension=0.5}{wc2,vc5}
\fmf{plain,tension=0.5}{vc3,vc4}
\fmf{phantom,tension=0.5}{vc4,wc1}
\fmf{plain,tension=0.5}{vc5,vc6}
\fmf{plain,tension=0.5}{vc6,vc7}
\fmf{plain,tension=0.5}{vc7,vc8}
\fmf{plain,tension=0.5,right=0,width=1mm}{v5,v10}
\fmffreeze
\fmf{dots,tension=0.5}{vc4,vc5}
\fmfposition
\fmfipath{p[]}
\fmfiset{p1}{vpath(__v5,__vc2)}
\fmfiset{p2}{vpath(__v10,__vc8)}
\fmfipair{wz[]}
\fmfiequ{wz2}{point length(p2)/2 of p2}
\vvertex{wz1}{wz2}{p1}
\svertex{wz2}{p2}
\wigglywrap{wz1}{v5}{v10}{wz2}
\fmfposition
\end{fmfchar*}}
\qquad
\fmfframe(3,1)(1,4){%
\begin{fmfchar*}(25,20)
\fmftop{v1}
\fmfbottom{v5}
\fmfforce{(0w,h)}{v1}
\fmfforce{(0w,0)}{v5}
\fmffixed{(0.2w,0)}{v1,v2}
\fmffixed{(0.2w,0)}{v2,v3}
\fmffixed{(0.2w,0)}{v3,w1}
\fmffixed{(0.2w,0)}{w1,v4}
\fmffixed{(0.2w,0)}{v4,v9}
\fmffixed{(0.2w,0)}{v5,v6}
\fmffixed{(0.2w,0)}{v6,v7}
\fmffixed{(0.2w,0)}{v7,w2}
\fmffixed{(0.2w,0)}{w2,v8}
\fmffixed{(0.2w,0)}{v8,v10}
\fmffixed{(0,whatever)}{vc1,vc2}
\fmffixed{(0,whatever)}{vc3,vc4}
\fmffixed{(0,whatever)}{vc5,vc6}
\fmffixed{(0,whatever)}{vc7,vc8}
\fmf{plain,tension=0.25,right=0.25}{v1,vc1}
\fmf{plain,tension=0.25,left=0.25}{v2,vc1}
\fmf{plain,left=0.25}{v5,vc2}
\fmf{plain,tension=1,left=0.25}{v3,vc3}
\fmf{phantom,tension=1,left=0.25}{w1,wc1}
\fmf{plain,tension=1,left=0.25}{v4,vc5}
\fmf{plain,tension=1,left=0.25}{v9,vc7}
\fmf{plain,left=0.25}{w2,vc6}
\fmf{plain,tension=0.25,left=0.25}{v8,vc8}
\fmf{plain,tension=0.25,right=0.25}{v10,vc8}
\fmf{plain,left=0.25}{v6,vc4}
\fmf{phantom,left=0.25}{v7,wc2}
\fmf{plain,tension=0.5}{vc1,vc2}
\fmf{phantom,tension=0.5}{wc1,wc2}
\fmf{plain,tension=0.5}{vc2,vc3}
\fmf{phantom,tension=0.5}{wc2,vc5}
\fmf{plain,tension=0.5}{vc3,vc4}
\fmf{phantom,tension=0.5}{vc4,wc1}
\fmf{plain,tension=0.5}{vc5,vc6}
\fmf{plain,tension=0.5}{vc6,vc7}
\fmf{plain,tension=0.5}{vc7,vc8}
\fmf{plain,tension=0.5,right=0,width=1mm}{v5,v10}
\fmffreeze
\fmf{dots,tension=0.5}{vc4,vc5}
\fmfposition
\fmfipath{p[]}
\fmfiset{p1}{vpath(__v5,__vc2)}
\fmfiset{p2}{vpath(__vc7,__vc8)}
\fmfipair{wz[]}
\svertex{wz1}{p1}
\svertex{wz2}{p2}
\wigglywrap{wz1}{v5}{v10}{wz2}
\fmfposition
\end{fmfchar*}}
\qquad
\fmfframe(3,1)(1,4){%
\begin{fmfchar*}(25,20)
\fmftop{v1}
\fmfbottom{v5}
\fmfforce{(0w,h)}{v1}
\fmfforce{(0w,0)}{v5}
\fmffixed{(0.2w,0)}{v1,v2}
\fmffixed{(0.2w,0)}{v2,v3}
\fmffixed{(0.2w,0)}{v3,w1}
\fmffixed{(0.2w,0)}{w1,v4}
\fmffixed{(0.2w,0)}{v4,v9}
\fmffixed{(0.2w,0)}{v5,v6}
\fmffixed{(0.2w,0)}{v6,v7}
\fmffixed{(0.2w,0)}{v7,w2}
\fmffixed{(0.2w,0)}{w2,v8}
\fmffixed{(0.2w,0)}{v8,v10}
\fmffixed{(0,whatever)}{vc1,vc2}
\fmffixed{(0,whatever)}{vc3,vc4}
\fmffixed{(0,whatever)}{vc5,vc6}
\fmffixed{(0,whatever)}{vc7,vc8}
\fmf{plain,tension=0.25,right=0.25}{v1,vc1}
\fmf{plain,tension=0.25,left=0.25}{v2,vc1}
\fmf{plain,left=0.25}{v5,vc2}
\fmf{plain,tension=1,left=0.25}{v3,vc3}
\fmf{phantom,tension=1,left=0.25}{w1,wc1}
\fmf{plain,tension=1,left=0.25}{v4,vc5}
\fmf{plain,tension=1,left=0.25}{v9,vc7}
\fmf{plain,left=0.25}{w2,vc6}
\fmf{plain,tension=0.25,left=0.25}{v8,vc8}
\fmf{plain,tension=0.25,right=0.25}{v10,vc8}
\fmf{plain,left=0.25}{v6,vc4}
\fmf{phantom,left=0.25}{v7,wc2}
\fmf{plain,tension=0.5}{vc1,vc2}
\fmf{phantom,tension=0.5}{wc1,wc2}
\fmf{plain,tension=0.5}{vc2,vc3}
\fmf{phantom,tension=0.5}{wc2,vc5}
\fmf{plain,tension=0.5}{vc3,vc4}
\fmf{phantom,tension=0.5}{vc4,wc1}
\fmf{plain,tension=0.5}{vc5,vc6}
\fmf{plain,tension=0.5}{vc6,vc7}
\fmf{plain,tension=0.5}{vc7,vc8}
\fmf{plain,tension=0.5,right=0,width=1mm}{v5,v10}
\fmffreeze
\fmf{dots,tension=0.5}{vc4,vc5}
\fmfposition
\fmfipath{p[]}
\fmfiset{p1}{vpath(__vc1,__vc2)}
\fmfiset{p2}{vpath(__v10,__vc8)}
\fmfipair{wz[]}
\svertex{wz1}{p1}
\svertex{wz2}{p2}
\wigglywrap{wz1}{v5}{v10}{wz2}
\fmfposition
\end{fmfchar*}}
\qquad
\fmfframe(3,1)(1,4){%
\begin{fmfchar*}(25,20)
\fmftop{v1}
\fmfbottom{v5}
\fmfforce{(0w,h)}{v1}
\fmfforce{(0w,0)}{v5}
\fmffixed{(0.2w,0)}{v1,v2}
\fmffixed{(0.2w,0)}{v2,v3}
\fmffixed{(0.2w,0)}{v3,w1}
\fmffixed{(0.2w,0)}{w1,v4}
\fmffixed{(0.2w,0)}{v4,v9}
\fmffixed{(0.2w,0)}{v5,v6}
\fmffixed{(0.2w,0)}{v6,v7}
\fmffixed{(0.2w,0)}{v7,w2}
\fmffixed{(0.2w,0)}{w2,v8}
\fmffixed{(0.2w,0)}{v8,v10}
\fmffixed{(0,whatever)}{vc1,vc2}
\fmffixed{(0,whatever)}{vc3,vc4}
\fmffixed{(0,whatever)}{vc5,vc6}
\fmffixed{(0,whatever)}{vc7,vc8}
\fmf{plain,tension=0.25,right=0.25}{v1,vc1}
\fmf{plain,tension=0.25,left=0.25}{v2,vc1}
\fmf{plain,left=0.25}{v5,vc2}
\fmf{plain,tension=1,left=0.25}{v3,vc3}
\fmf{phantom,tension=1,left=0.25}{w1,wc1}
\fmf{plain,tension=1,left=0.25}{v4,vc5}
\fmf{plain,tension=1,left=0.25}{v9,vc7}
\fmf{plain,left=0.25}{w2,vc6}
\fmf{plain,tension=0.25,left=0.25}{v8,vc8}
\fmf{plain,tension=0.25,right=0.25}{v10,vc8}
\fmf{plain,left=0.25}{v6,vc4}
\fmf{phantom,left=0.25}{v7,wc2}
\fmf{plain,tension=0.5}{vc1,vc2}
\fmf{phantom,tension=0.5}{wc1,wc2}
\fmf{plain,tension=0.5}{vc2,vc3}
\fmf{phantom,tension=0.5}{wc2,vc5}
\fmf{plain,tension=0.5}{vc3,vc4}
\fmf{phantom,tension=0.5}{vc4,wc1}
\fmf{plain,tension=0.5}{vc5,vc6}
\fmf{plain,tension=0.5}{vc6,vc7}
\fmf{plain,tension=0.5}{vc7,vc8}
\fmf{plain,tension=0.5,right=0,width=1mm}{v5,v10}
\fmffreeze
\fmf{dots,tension=0.5}{vc4,vc5}
\fmfposition
\fmfipath{p[]}
\fmfiset{p1}{vpath(__vc1,__vc2)}
\fmfiset{p2}{vpath(__vc7,__vc8)}
\fmfipair{wz[]}
\svertex{wz1}{p1}
\svertex{wz2}{p2}
\wigglywrap{wz1}{v5}{v10}{wz2}
\fmfposition
\end{fmfchar*}}}
}
\\
\vspace{0.2cm}
$\Huge{\vdots}$
\vspace{0.3cm}
\\
\subfigure[$W_{L,j}\quad 0<j<L-1$]{
\fmfframe(3,1)(1,4){%
\begin{fmfchar*}(30,20)
\fmftop{v1}
\fmfbottom{v5}
\fmfforce{(0w,h)}{v1}
\fmfforce{(0w,0)}{v5}
\fmffixed{(0.25w,0)}{v1,v2}
\fmffixed{(0.16w,0)}{v2,v3}
\fmffixed{(0.16w,0)}{v3,w1}
\fmffixed{(0.16w,0)}{w1,v4}
\fmffixed{(0.16w,0)}{v4,v9}
\fmffixed{(0.25w,0)}{v5,v6}
\fmffixed{(0.16w,0)}{v6,v7}
\fmffixed{(0.16w,0)}{v7,w2}
\fmffixed{(0.16w,0)}{w2,v8}
\fmffixed{(0.16w,0)}{v8,v10}
\fmffixed{(0,whatever)}{vc1,vc2}
\fmffixed{(0,whatever)}{vc3,vc4}
\fmffixed{(0,whatever)}{vc5,vc6}
\fmffixed{(0,whatever)}{vc7,vc8}
\fmf{phantom,tension=0.25,right=0.25}{v2,vc1}
\fmf{phantom,tension=0.25,left=0.25}{v3,vc1}
\fmf{phantom,left=0.25}{v6,vc2}
\fmf{plain,tension=1,left=0.25}{w1,vc3}
\fmf{phantom,tension=1,left=0.25}{v4,wc1}
\fmf{plain,tension=1,left=0.25}{v4,vc5}
\fmf{plain,tension=1,left=0.25}{v9,vc7}
\fmf{plain,left=0.25}{w2,vc6}
\fmf{plain,tension=0.25,left=0.25}{v8,vc8}
\fmf{plain,tension=0.25,right=0.25}{v10,vc8}
\fmf{plain,left=0.25}{v7,vc4}
\fmf{phantom,left=0.25}{w2,wc2}
\fmf{phantom,tension=0.5}{vc1,vc2}
\fmf{phantom,tension=0.5}{vc2,vc3}
\fmf{phantom,tension=0.5}{wc2,vc5}
\fmf{plain,tension=0.5}{vc3,vc4}
\fmf{phantom,tension=0.5}{vc4,wc1}
\fmf{plain,tension=0.5}{vc5,vc6}
\fmf{plain,tension=0.5}{vc6,vc7}
\fmf{plain,tension=0.5}{vc7,vc8}
\fmf{plain}{v1,v5}
\fmf{plain,tension=0.5,right=0,width=1mm}{v5,v10}
\fmffreeze
\fmffixed{(0.32w,0)}{v1,v11}
\fmffixed{(0.32w,0)}{v5,v12}
\fmf{plain,tension=0.25,right=0.25}{v3,vc3}
\fmf{plain}{v11,v12}
\fmf{dots,tension=0.5}{vc4,vc5}
\fmfposition
\fmfipath{p[]}
\fmfipair{wz[]}
\fmfiset{p1}{vpath(__v7,__vc4)}
\fmfiset{p2}{vpath(__v1,__v5)}
\fmfiset{p3}{vpath(__v10,__vc8)}
\fmfiset{p4}{vpath(__v11,__v12)}
\svertex{wz1}{p1}
\vvertex{wz2}{wz1}{p2}
\svertex{wz3}{p2}
\svertex{wz4}{p3}
\svertex{wz5}{p4}
\fmfiequ{wz6}{(xpart(wz3)+6,ypart(wz3))}
\fmfiequ{wz7}{(xpart(wz5)-6,ypart(wz5))}
\fmfi{wiggly}{wz5..wz1}
\fmfi{wiggly}{wz6..wz3}
\fmfi{wiggly}{wz7..wz5}
\fmfi{dots}{wz6..wz7}
\wigglywrap{wz3}{v5}{v10}{wz4}
\end{fmfchar*}}
\qquad
\fmfframe(3,1)(1,4){%
\begin{fmfchar*}(30,20)
\fmftop{v1}
\fmfbottom{v5}
\fmfforce{(0w,h)}{v1}
\fmfforce{(0w,0)}{v5}
\fmffixed{(0.25w,0)}{v1,v2}
\fmffixed{(0.16w,0)}{v2,v3}
\fmffixed{(0.16w,0)}{v3,w1}
\fmffixed{(0.16w,0)}{w1,v4}
\fmffixed{(0.16w,0)}{v4,v9}
\fmffixed{(0.25w,0)}{v5,v6}
\fmffixed{(0.16w,0)}{v6,v7}
\fmffixed{(0.16w,0)}{v7,w2}
\fmffixed{(0.16w,0)}{w2,v8}
\fmffixed{(0.16w,0)}{v8,v10}
\fmffixed{(0,whatever)}{vc1,vc2}
\fmffixed{(0,whatever)}{vc3,vc4}
\fmffixed{(0,whatever)}{vc5,vc6}
\fmffixed{(0,whatever)}{vc7,vc8}
\fmf{phantom,tension=0.25,right=0.25}{v2,vc1}
\fmf{phantom,tension=0.25,left=0.25}{v3,vc1}
\fmf{phantom,left=0.25}{v6,vc2}
\fmf{plain,tension=1,left=0.25}{w1,vc3}
\fmf{phantom,tension=1,left=0.25}{v4,wc1}
\fmf{plain,tension=1,left=0.25}{v4,vc5}
\fmf{plain,tension=1,left=0.25}{v9,vc7}
\fmf{plain,left=0.25}{w2,vc6}
\fmf{plain,tension=0.25,left=0.25}{v8,vc8}
\fmf{plain,tension=0.25,right=0.25}{v10,vc8}
\fmf{plain,left=0.25}{v7,vc4}
\fmf{phantom,left=0.25}{w2,wc2}
\fmf{phantom,tension=0.5}{vc1,vc2}
\fmf{phantom,tension=0.5}{vc2,vc3}
\fmf{phantom,tension=0.5}{wc2,vc5}
\fmf{plain,tension=0.5}{vc3,vc4}
\fmf{phantom,tension=0.5}{vc4,wc1}
\fmf{plain,tension=0.5}{vc5,vc6}
\fmf{plain,tension=0.5}{vc6,vc7}
\fmf{plain,tension=0.5}{vc7,vc8}
\fmf{plain}{v1,v5}
\fmf{plain,tension=0.5,right=0,width=1mm}{v5,v10}
\fmffreeze
\fmffixed{(0.32w,0)}{v1,v11}
\fmffixed{(0.32w,0)}{v5,v12}
\fmf{plain,tension=0.25,right=0.25}{v3,vc3}
\fmf{plain}{v11,v12}
\fmf{dots,tension=0.5}{vc4,vc5}
\fmfposition
\fmfipath{p[]}
\fmfipair{wz[]}
\fmfiset{p1}{vpath(__v7,__vc4)}
\fmfiset{p2}{vpath(__v1,__v5)}
\fmfiset{p3}{vpath(__vc7,__vc8)}
\fmfiset{p4}{vpath(__v11,__v12)}
\svertex{wz1}{p1}
\vvertex{wz2}{wz1}{p2}
\svertex{wz3}{p2}
\svertex{wz4}{p3}
\svertex{wz5}{p4}
\fmfiequ{wz6}{(xpart(wz3)+6,ypart(wz3))}
\fmfiequ{wz7}{(xpart(wz5)-6,ypart(wz5))}
\fmfi{wiggly}{wz5..wz1}
\fmfi{wiggly}{wz6..wz3}
\fmfi{wiggly}{wz7..wz5}
\fmfi{dots}{wz6..wz7}
\wigglywrap{wz3}{v5}{v10}{wz4}
\end{fmfchar*}}
\qquad
\fmfframe(3,1)(1,4){%
\begin{fmfchar*}(30,20)
\fmftop{v1}
\fmfbottom{v5}
\fmfforce{(0w,h)}{v1}
\fmfforce{(0w,0)}{v5}
\fmffixed{(0.25w,0)}{v1,v2}
\fmffixed{(0.16w,0)}{v2,v3}
\fmffixed{(0.16w,0)}{v3,w1}
\fmffixed{(0.16w,0)}{w1,v4}
\fmffixed{(0.16w,0)}{v4,v9}
\fmffixed{(0.25w,0)}{v5,v6}
\fmffixed{(0.16w,0)}{v6,v7}
\fmffixed{(0.16w,0)}{v7,w2}
\fmffixed{(0.16w,0)}{w2,v8}
\fmffixed{(0.16w,0)}{v8,v10}
\fmffixed{(0,whatever)}{vc1,vc2}
\fmffixed{(0,whatever)}{vc3,vc4}
\fmffixed{(0,whatever)}{vc5,vc6}
\fmffixed{(0,whatever)}{vc7,vc8}
\fmf{phantom,tension=0.25,right=0.25}{v2,vc1}
\fmf{phantom,tension=0.25,left=0.25}{v3,vc1}
\fmf{phantom,left=0.25}{v6,vc2}
\fmf{plain,tension=1,left=0.25}{w1,vc3}
\fmf{phantom,tension=1,left=0.25}{v4,wc1}
\fmf{plain,tension=1,left=0.25}{v4,vc5}
\fmf{plain,tension=1,left=0.25}{v9,vc7}
\fmf{plain,left=0.25}{w2,vc6}
\fmf{plain,tension=0.25,left=0.25}{v8,vc8}
\fmf{plain,tension=0.25,right=0.25}{v10,vc8}
\fmf{plain,left=0.25}{v7,vc4}
\fmf{phantom,left=0.25}{w2,wc2}
\fmf{phantom,tension=0.5}{vc1,vc2}
\fmf{phantom,tension=0.5}{vc2,vc3}
\fmf{phantom,tension=0.5}{wc2,vc5}
\fmf{plain,tension=0.5}{vc3,vc4}
\fmf{phantom,tension=0.5}{vc4,wc1}
\fmf{plain,tension=0.5}{vc5,vc6}
\fmf{plain,tension=0.5}{vc6,vc7}
\fmf{plain,tension=0.5}{vc7,vc8}
\fmf{plain}{v1,v5}
\fmf{plain,tension=0.5,right=0,width=1mm}{v5,v10}
\fmffreeze
\fmffixed{(0.32w,0)}{v1,v11}
\fmffixed{(0.32w,0)}{v5,v12}
\fmf{plain,tension=0.25,right=0.25}{v3,vc3}
\fmf{plain}{v11,v12}
\fmf{dots,tension=0.5}{vc4,vc5}
\fmfposition
\fmfipath{p[]}
\fmfipair{wz[]}
\fmfiset{p1}{vpath(__vc3,__vc4)}
\fmfiset{p2}{vpath(__v1,__v5)}
\fmfiset{p3}{vpath(__v10,__vc8)}
\fmfiset{p4}{vpath(__v11,__v12)}
\svertex{wz1}{p1}
\vvertex{wz2}{wz1}{p2}
\svertex{wz3}{p2}
\svertex{wz4}{p3}
\svertex{wz5}{p4}
\fmfiequ{wz6}{(xpart(wz3)+6,ypart(wz3))}
\fmfiequ{wz7}{(xpart(wz5)-6,ypart(wz5))}
\fmfi{wiggly}{wz5..wz1}
\fmfi{wiggly}{wz6..wz3}
\fmfi{wiggly}{wz7..wz5}
\fmfi{dots}{wz6..wz7}
\wigglywrap{wz3}{v5}{v10}{wz4}
\end{fmfchar*}}
\qquad
\fmfframe(3,1)(1,4){%
\begin{fmfchar*}(30,20)
\fmftop{v1}
\fmfbottom{v5}
\fmfforce{(0w,h)}{v1}
\fmfforce{(0w,0)}{v5}
\fmffixed{(0.25w,0)}{v1,v2}
\fmffixed{(0.16w,0)}{v2,v3}
\fmffixed{(0.16w,0)}{v3,w1}
\fmffixed{(0.16w,0)}{w1,v4}
\fmffixed{(0.16w,0)}{v4,v9}
\fmffixed{(0.25w,0)}{v5,v6}
\fmffixed{(0.16w,0)}{v6,v7}
\fmffixed{(0.16w,0)}{v7,w2}
\fmffixed{(0.16w,0)}{w2,v8}
\fmffixed{(0.16w,0)}{v8,v10}
\fmffixed{(0,whatever)}{vc1,vc2}
\fmffixed{(0,whatever)}{vc3,vc4}
\fmffixed{(0,whatever)}{vc5,vc6}
\fmffixed{(0,whatever)}{vc7,vc8}
\fmf{phantom,tension=0.25,right=0.25}{v2,vc1}
\fmf{phantom,tension=0.25,left=0.25}{v3,vc1}
\fmf{phantom,left=0.25}{v6,vc2}
\fmf{plain,tension=1,left=0.25}{w1,vc3}
\fmf{phantom,tension=1,left=0.25}{v4,wc1}
\fmf{plain,tension=1,left=0.25}{v4,vc5}
\fmf{plain,tension=1,left=0.25}{v9,vc7}
\fmf{plain,left=0.25}{w2,vc6}
\fmf{plain,tension=0.25,left=0.25}{v8,vc8}
\fmf{plain,tension=0.25,right=0.25}{v10,vc8}
\fmf{plain,left=0.25}{v7,vc4}
\fmf{phantom,left=0.25}{w2,wc2}
\fmf{phantom,tension=0.5}{vc1,vc2}
\fmf{phantom,tension=0.5}{vc2,vc3}
\fmf{phantom,tension=0.5}{wc2,vc5}
\fmf{plain,tension=0.5}{vc3,vc4}
\fmf{phantom,tension=0.5}{vc4,wc1}
\fmf{plain,tension=0.5}{vc5,vc6}
\fmf{plain,tension=0.5}{vc6,vc7}
\fmf{plain,tension=0.5}{vc7,vc8}
\fmf{plain}{v1,v5}
\fmf{plain,tension=0.5,right=0,width=1mm}{v5,v10}
\fmffreeze
\fmffixed{(0.32w,0)}{v1,v11}
\fmffixed{(0.32w,0)}{v5,v12}
\fmf{plain,tension=0.25,right=0.25}{v3,vc3}
\fmf{plain}{v11,v12}
\fmf{dots,tension=0.5}{vc4,vc5}
\fmfposition
\fmfipath{p[]}
\fmfipair{wz[]}
\fmfiset{p1}{vpath(__vc3,__vc4)}
\fmfiset{p2}{vpath(__v1,__v5)}
\fmfiset{p3}{vpath(__vc7,__vc8)}
\fmfiset{p4}{vpath(__v11,__v12)}
\svertex{wz1}{p1}
\vvertex{wz2}{wz1}{p2}
\svertex{wz3}{p2}
\svertex{wz4}{p3}
\svertex{wz5}{p4}
\fmfiequ{wz6}{(xpart(wz3)+6,ypart(wz3))}
\fmfiequ{wz7}{(xpart(wz5)-6,ypart(wz5))}
\fmfi{wiggly}{wz5..wz1}
\fmfi{wiggly}{wz6..wz3}
\fmfi{wiggly}{wz7..wz5}
\fmfi{dots}{wz6..wz7}
\wigglywrap{wz3}{v5}{v10}{wz4}
\end{fmfchar*}}
}
\\
\vspace{0.2cm}
$\Huge{\vdots}$
\vspace{0.3cm}
\\
\subfigure[$W_{L,L-1}$]
{
\fmfframe(3,1)(1,4){%
\begin{fmfchar*}(25,20)
\fmftop{v1}
\fmfbottom{v5}
\fmfforce{(0w,h)}{v1}
\fmfforce{(0w,0)}{v5}
\fmffixed{(0.2w,0)}{v1,v2}
\fmffixed{(0.2w,0)}{v2,v3}
\fmffixed{(0.2w,0)}{v3,w1}
\fmffixed{(0.2w,0)}{w1,v4}
\fmffixed{(0.2w,0)}{v4,v9}
\fmffixed{(0.2w,0)}{v5,v6}
\fmffixed{(0.2w,0)}{v6,v7}
\fmffixed{(0.2w,0)}{v7,w2}
\fmffixed{(0.2w,0)}{w2,v8}
\fmffixed{(0.2w,0)}{v8,v10}
\fmffixed{(0,whatever)}{vc7,vc8}
\fmf{plain,tension=0.25,right=0.25}{v4,vc7}
\fmf{plain,tension=0.25,left=0.25}{v9,vc7}
\fmf{plain,tension=0.25,left=0.25}{v8,vc8}
\fmf{plain,tension=0.25,right=0.25}{v10,vc8}
\fmf{plain}{v2,v6}
\fmf{plain}{w2,w1}
\fmf{plain,tension=0.5}{vc7,vc8}
\fmf{plain}{v1,v5}
\fmf{plain,tension=0.5,right=0,width=1mm}{v5,v10}
\fmffreeze
\fmfposition
\fmfipath{p[]}
\fmfipair{wz[]}
\fmfiset{p1}{vpath(__v2,__v6)}
\fmfiset{p2}{vpath(__w1,__w2)}
\fmfiset{p3}{vpath(__v10,__vc8)}
\fmfiset{p4}{vpath(__v1,__v5)}
\fmfiset{p5}{vpath(__vc8,__vc7)}
\svertex{wz1}{p1}
\svertex{wz2}{p2}
\svertex{wz3}{p3}
\svertex{wz4}{p4}
\svertex{wz5}{p5}
\fmfiequ{wz6}{(xpart(wz1)+6,ypart(wz1))}
\fmfiequ{wz7}{(xpart(wz2)-6,ypart(wz2))}
\fmfi{wiggly}{wz2..wz5}
\fmfi{dots}{wz6..wz7}
\fmfi{wiggly}{wz4..wz1}
\fmfi{wiggly}{wz1..wz6}
\fmfi{wiggly}{wz7..wz2}
\wigglywrap{wz4}{v5}{v10}{wz3}
\end{fmfchar*}}
\qquad
\fmfframe(3,1)(1,4){%
\begin{fmfchar*}(25,20)
\fmftop{v1}
\fmfbottom{v5}
\fmfforce{(0w,h)}{v1}
\fmfforce{(0w,0)}{v5}
\fmffixed{(0.2w,0)}{v1,v2}
\fmffixed{(0.2w,0)}{v2,v3}
\fmffixed{(0.2w,0)}{v3,w1}
\fmffixed{(0.2w,0)}{w1,v4}
\fmffixed{(0.2w,0)}{v4,v9}
\fmffixed{(0.2w,0)}{v5,v6}
\fmffixed{(0.2w,0)}{v6,v7}
\fmffixed{(0.2w,0)}{v7,w2}
\fmffixed{(0.2w,0)}{w2,v8}
\fmffixed{(0.2w,0)}{v8,v10}
\fmffixed{(0,whatever)}{vc7,vc8}
\fmf{plain,tension=0.25,right=0.25}{v4,vc7}
\fmf{plain,tension=0.25,left=0.25}{v9,vc7}
\fmf{plain,tension=0.25,left=0.25}{v8,vc8}
\fmf{plain,tension=0.25,right=0.25}{v10,vc8}
\fmf{plain}{v2,v6}
\fmf{plain}{w2,w1}
\fmf{plain,tension=0.5}{vc7,vc8}
\fmf{plain}{v1,v5}
\fmf{plain,tension=0.5,right=0,width=1mm}{v5,v10}
\fmffreeze
\fmfposition
\fmfipath{p[]}
\fmfipair{wz[]}
\fmfiset{p1}{vpath(__v2,__v6)}
\fmfiset{p2}{vpath(__w1,__w2)}
\fmfiset{p3}{vpath(__v10,__vc8)}
\fmfiset{p4}{vpath(__v1,__v5)}
\fmfiset{p5}{vpath(__vc8,__vc7)}
\svertex{wz1}{p1}
\svertex{wz2}{p2}
\fmfiequ{wz3}{point 2*length(p5)/3 of p5}
\svertex{wz4}{p4}
\fmfiequ{wz5}{point length(p5)/3 of p5}
\fmfiequ{wz6}{(xpart(wz1)+6,ypart(wz1))}
\fmfiequ{wz7}{(xpart(wz2)-6,ypart(wz2))}
\fmfi{wiggly}{wz2..wz5}
\fmfi{dots}{wz6..wz7}
\fmfi{wiggly}{wz4..wz1}
\fmfi{wiggly}{wz1..wz6}
\fmfi{wiggly}{wz7..wz2}
\wigglywrap{wz4}{v5}{v10}{wz3}
\end{fmfchar*}}
}
\caption{Relevant diagrams after cancellations}
\label{graphs}
\end{figure}

We conclude that the only relevant diagrams with vectors are the ones in which the vectors interact with single Z lines through the $V_2$ vertex. In each set the number of  contributing graphs is now reduced to at most four, as depicted in Fig.~\ref{graphs}.

The D-algebra can be performed straightforwardly and it is easy to realize that the sum of the graphs with $j$ vectors produces the same momentum integrals as the sum of the diagrams with $L-j-1$ vectors.

\begin{figure}[h]
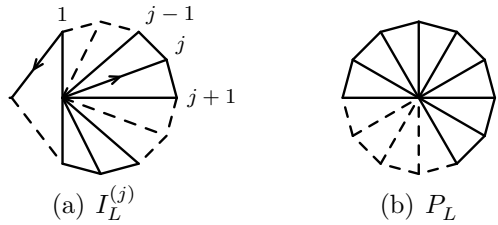

\unitlength=0.75mm
\settoheight{\eqoff}{$\times$}%
\setlength{\eqoff}{0.5\eqoff}%
\addtolength{\eqoff}{-14.5\unitlength}%
\settoheight{\eqofftwo}{$\times$}%
\setlength{\eqofftwo}{0.5\eqofftwo}%
\addtolength{\eqofftwo}{-7.5\unitlength}%
\centering
\subfigure[$I_L^{(j)}$]{
\label{ILk}
\raisebox{\eqoff}{%
\fmfframe(0,0)(0,0){%
\begin{fmfchar*}(30,30)
  \fmfleft{in}
  \fmfright{out}
  \fmf{plain,tension=1}{in,vi}
  \fmf{phantom,tension=1}{out,v7}
\fmfpoly{phantom}{v12,v11,v10,v9,v8,v7,v6,v5,v4,v3,v2,v1}
\fmffixed{(0.9w,0)}{v1,v7}
\fmffixed{(0.3w,0)}{vi,v0}
\fmffixed{(0,whatever)}{v0,v3}
\fmf{dashes}{v3,v4}
\fmf{dashes}{v4,v5}
\fmf{plain}{v5,v6}
\fmf{plain}{v6,v7}
\fmf{dashes}{v7,v8}
\fmf{dashes}{v8,v9}
\fmf{plain}{v9,v10}
\fmf{plain}{v10,v11}
\fmf{derplain}{v3,vi}
\fmf{dashes}{vi,v11}
\fmf{plain}{v0,v3}
\fmf{dashes}{v0,v4}
\fmf{plain}{v0,v5}
\fmf{derplain}{v0,v6}
\fmf{plain}{v0,v7}
\fmf{dashes}{v0,v8}
\fmf{plain}{v0,v9}
\fmf{plain}{v0,v10}
\fmf{plain}{v0,v11}
\fmfposition
\fmfipair{w[]}
\fmfiequ{w1}{(xpart(vloc(__v3)),ypart(vloc(__v3)))}
\fmfiequ{w2}{(xpart(vloc(__v4)),ypart(vloc(__v4)))}
\fmfiequ{w3}{(xpart(vloc(__v5)),ypart(vloc(__v5)))}
\fmfiequ{w4}{(xpart(vloc(__v6)),ypart(vloc(__v6)))}
\fmfiequ{w5}{(xpart(vloc(__v7)),ypart(vloc(__v7)))}
\fmfiequ{w6}{(xpart(vloc(__v8)),ypart(vloc(__v8)))}
\fmfiequ{w7}{(xpart(vloc(__v9)),ypart(vloc(__v9)))}
\fmfiv{l=\scriptsize{$1$},l.a=90,l.d=4}{w1}
\fmfiv{l=\scriptsize{$j-1$},l.a=45,l.d=4}{w3}
\fmfiv{l=\scriptsize{$j$},l.a=30,l.d=4}{w4}
\fmfiv{l=\scriptsize{$j+1$},l.a=0,l.d=4}{w5}
\end{fmfchar*}}}
}
\qquad\qquad
\subfigure[$P_L^{\protect\phantom{(}}$]{
\label{PL}
\raisebox{\eqoff}{%
\fmfframe(0,0)(0,0){%
\begin{fmfchar*}(30,30)
  \fmfleft{in}
  \fmfright{out}
  \fmf{phantom,tension=1}{in,v1}
  \fmf{phantom,tension=1}{out,v7}
\fmfpoly{phantom}{v12,v11,v10,v9,v8,v7,v6,v5,v4,v3,v2,v1}
\fmffixed{(0.9w,0)}{v1,v7}
\fmfforce{(0.5w,0.5h)}{v0}
\fmfposition
\fmffreeze
\fmf{plain}{v1,v2}
\fmf{plain}{v2,v3}
\fmf{plain}{v3,v4}
\fmf{plain}{v4,v5}
\fmf{plain}{v5,v6}
\fmf{plain}{v6,v7}
\fmf{plain}{v7,v8}
\fmf{plain}{v8,v9}
\fmf{dashes}{v9,v10}
\fmf{dashes}{v10,v11}
\fmf{dashes}{v11,v12}
\fmf{dashes}{v12,v1}
\fmf{plain}{v1,v0}
\fmf{plain}{v2,v0}
\fmf{plain}{v3,v0}
\fmf{plain}{v4,v0}
\fmf{plain}{v5,v0}
\fmf{plain}{v6,v0}
\fmf{plain}{v7,v0}
\fmf{plain}{v8,v0}
\fmf{plain}{v9,v0}
\fmf{dashes}{v10,v0}
\fmf{dashes}{v11,v0}
\fmf{dashes}{v12,v0}
\end{fmfchar*}}}
}
\caption{$L$-loop momentum integrals}
\label{integrals}
\end{figure}

We denote the color factors for the various chiral structures by $C_{L,j}$, where the subscript $j$ stands for the number of vectors entering the diagrams. Then we have
\beq
C_{L,j}=(q-\bar{q})^2\left[q^{2(L-j-1)}+\bar{q}^{2(L-j-1)}\right]=-8\sin^2(\pi\beta)\cos[2\pi\beta(L-j-1)]
\label{colfact}
\eeq
for $j=0,\dots,L-1$.\\
Using these color factors, we can write the contributions from each class:
\begin{equation}
\begin{aligned}
W_{L,0}-S_L&=(g^2 N)^L\ C_{L,0}(K_L-J_L)\col\\
\vdots\\
W_{L,j}&=2(g^2 N)^L\ C_{L,j}I_L^{(j+1)}\col\\
\vdots\\
W_{L,L-1}&=-(g^2 N)^L\ C_{L,L-1}(K_L-J_L)\col
\end{aligned}
\end{equation}
where the integrals $I_L^{(j)}$ are shown in Fig.~\ref{ILk}.\\
These integrals satisfy the relation
\begin{equation}
\label{ILrel}
I_L^{(j)}=-I_L^{(L-j+1)}
\pnt
\end{equation}
The combination $(K_L-J_L)$, which is relevant for $(W_{L,0}-S_L)$ and for class $W_{L,L-1}$, can be written in terms of $I_L^{(1)}$ and of the integral $P_L$ (shown in Fig.~\ref{PL}) as
\begin{equation}
K_L-J_L=P_L-2I_L^{(1)}
\pnt
\end{equation}

We can now collect all these results and obtain the correct value for the anomalous dimension:
\begin{equation}
\gamma_L(\mathcal{O}_{1,L})=\gamma_L^{as}+\delta\gamma_L(\mathcal{O}_{1,L})
\pnt
\end{equation}
Since both the $P_L$ and the $I_L^{(j)}$ are free of subdivergences, their Laurent expansion in $\varepsilon$ will present only poles of the first order. Thus we can write 
\begin{equation}
\delta\gamma_L(\mathcal{O}_{1,L})=-2L(g^2 N)^L\lim_{\varepsilon\rightarrow0}\varepsilon\Bigg[(C_{L,0}-C_{L,L-1})P_L(\varepsilon)-2\sum_{j=0}^{[\frac{L}{2}]-1}(C_{L,j}-C_{L,L-j-1})I_L^{(j+1)}(\varepsilon)\Bigg]
\pnt
\end{equation}

\section{Computation of the integrals}
In order to obtain the actual value of the anomalous dimension for a given loop order $L$, we need the explicit values of the coefficients of the $1/\varepsilon$ poles in the expansions of the momentum integrals $P_L$ and $I_L^{(j)}$.

For the integrals $P_L$, this result is known as a function of $L$~\cite{Broadhurst:1985vq,Usyukina:1991cp}:
\begin{equation}
P_L\sim\frac{1}{\varepsilon}\frac{1}{(4\pi)^{2L}}\frac{2}{L}\binom{2L-3}{L-1}\z(2L-3)
\col
\end{equation}
where the symbol $\sim$ means that we are only interested in the divergent part.

We were not able to find a general formula for the divergent parts of the integrals $I_L^{(j)}$ as functions of $L$ and $j$. However, it is possible to find recurrence relations which allow to compute the required integrals for \textit{any} fixed value of $L$ and $j$.

These recurrence relations can be found applying the technique of integration by parts, as described in~\cite{Broadhurst:1985vq}, where integrals with the same topology, but without scalar products of momenta in the numerators, were considered. In order to follow this approach, we had to generalize the triangle rule of~\cite{Broadhurst:1985vq} to the case of lines with momenta in the numerators. The derivation of these generalized rules is shown in Appendix~\ref{app:rules}.\\
Using the scalar rule~\eqref{scalar-triangle}, any $I_L^{(j)}$ can be written in terms of a reduced set of integrals with at most seven loops and with generic propagator weights for the lines coming out from the operator insertion. An example of this procedure is presented in Appendix~\ref{app:example-rec}.\\
The fundamental integrals can be computed explicitly using the generalized rules. 

For the integrals $I_L^{(1)}$, also the technique described in~\cite{Usyukina:1991cp} can be used. This allowed us to guess a general expression for $I_L^{(1)}$ as a function of $L$:
\begin{equation}
\begin{aligned}
I_L^{(1)}&=\frac{1}{2}P_L+\frac{1}{L}\sum_{k=3}^{L-1}\binom{h(L,k)}{L-k}\zeta(h(L,k)+1)\\
&=\frac{1}{2}P_L+\frac{1}{L}\sum_{k=L-1-[\frac{L-1}{2}]}^{L-3}\binom{2k+1}{2k+3-L}\zeta(2k+1)+\frac{1}{2L}[1+(-1)^L](L-2)\zeta(L-1)
\col
\end{aligned}
\end{equation}
where $h(L,k)=2(L-k-1+[k/2])$. We could not apply the approach of~\cite{Usyukina:1991cp} to the other classes of integrals because we were not able to compute the needed higher-loop integrals with generic propagator weights.
For $L\ge 10$
it also takes too long to extract the pole part from the results which were 
obtained from the the recurrence formula. To find the results up to $L=11$ 
we have therefore employed GPXT \cite{Chetyrkin:1980pr} 
as described in Appendix~\ref{app:GPXT}.

Looking for a simple deformation of the formula for $I_L^{(1)}$, we were able to guess the general expression for $I_L^{(2)}$:
\begin{equation}
\begin{aligned}
I_L^{(2)}\vert_{L=2m}&=\frac{1}{2}P_L-\frac{1}{L}\sum_{k=L-1-[\frac{L-1}{2}]}^{L-3}{\left[\frac{2L}{L-1}(L-2-k)-1\right]}\binom{2k+1}{2k+3-L}\zeta(2k+1)\\
&\quad-\frac{1}{L}(L-2)(L-1)\zeta(L-1)\col\\
& \\
I_L^{(2)}\vert_{L=2m+1}&=\frac{1}{2}P_L-\frac{1}{L}\sum_{k=L-[\frac{L-1}{2}]}^{L-3}{\left[\frac{2L}{L-1}(L-2-k)-1\right]}\binom{2k+1}{2k+3-L}\zeta(2k+1)\\
&\quad-\frac{1}{2}(L-3)(L-1)\zeta(L)
\pnt
\end{aligned}
\end{equation}
Both these formulae have been verified up to $L=11$.

\section{Concluding remarks}
We explicitly computed all the relevant $I_L^{(j)}$ up to $L=11$. The corresponding results are shown in Appendix~\ref{app:results}.\\
First of all, we see that in all the cases we considered, wrapping interactions and range-$(L+1)$ subtractions only produce transcendental contributions.\\
Moreover, a very precise transcendentality pattern appears, as can be seen from Table~\ref{tab:zetas}: for every value of $L$, the term with maximum degree of transcendentality is always proportional to $\zeta(2L-3)$. Then a given number of terms with consecutive, lower odd degrees is present. This number is increased by one every two loops.\\
This particular behaviour is also confirmed by the general expressions we guessed for a subset of the relevant classes of integrals.\\
Using the recurrence relations obtained from the triangle rules, one can in principle compute the exact anomalous dimension for any single-impurity, length-$L$ operator at the critical loop order $L$.
\begin{table}
\begin{center}
\begin{tabular}{|c|c|}
\hline
$L$ & Transcendental terms \\
\hline
$L=4$ & $\zeta(3)$, $\zeta(5)$\\
$L=5$ & $\zeta(5)$, $\zeta(7)$\\
$L=6$ & $\zeta(5)$, $\zeta(7)$, $\zeta(9)$\\
$L=7$ & $\zeta(7)$, $\zeta(9)$, $\zeta(11)$\\
$L=8$ & $\zeta(7)$, $\zeta(9)$, $\zeta(11)$, $\zeta(13)$\\
$L=9$ & $\zeta(9)$, $\zeta(11)$, $\zeta(13)$, $\zeta(15)$\\
$L=10$ & $\zeta(9)$, $\zeta(11)$, $\zeta(13)$, $\zeta(15)$, $\zeta(17)$\\
$L=11$ & $\zeta(11)$, $\zeta(13)$, $\zeta(15)$, $\zeta(17)$, $\zeta(19)$\\
\hline
\end{tabular}
\end{center}
\caption{Transcendental terms produced for different values of $L$}
\label{tab:zetas}
\end{table}

\section*{Acknowledgements}
\noindent This work has been supported in
part by INFN, by the European Commission RTN
program MRTN--CT--2004--005104 and by the Italian MIUR-PRIN contract 
20075ATT78.

\newpage

\appendix

\section{Details on the computation of the integrals}
\subsection{Triangle rules}
\label{app:rules}
In this section we present the generalization of the triangle rule of~\cite{Broadhurst:1985vq} to the case of lines with momenta in the numerators, and show how they can be used to obtain recurrence relations for the integrals we need.\\
All the rules we need can be obtained from the following integration by parts identity, which is valid for $\alpha+\beta+1-D/2-\mathrm{dim}(f)/2>0$:
\begin{equation}
\begin{aligned}
0 &=\int\mathrm{d}^D l\frac{\partial}{\partial l^{\mu}}\frac{f(l)l^{\mu}}{(l+p_1)^{2\alpha}(l+p_2)^{2\beta}l^2}\\
&=\int\mathrm{d}^D l\Big(\partial_{\mu}f(l)l^{\mu}+D f(l)\\
&\qquad-2f(l)l^{\mu}\Big(\alpha\frac{(l+p_1)_{\mu}}{(l+p_1)^{2}}+\beta\frac{(l+p_2)_{\mu}}{(l+p_2)^2}+\frac{l^{\mu}}{l^2}\Big)\Big)\frac{1}{(l+p_1)^{2\alpha}(l+p_2)^{2\beta}l^2}\\
&=\int\mathrm{d}^D l\Big(\partial_{\mu}f(l)l^{\mu}\\
&\qquad-f(l)\Big(\alpha\frac{l^2+(l+p_1)^2-p_1^2}{(l+p1)^2}+\beta\frac{l^2+(l+p_2)^2-p_2^2}{(l+p_2)^2}-D+2\Big)\Big)\frac{1}{(l+p_1)^{2\alpha}(l+p_2)^{2\beta}l^2}\\
&=\int\mathrm{d}^D l\Big(\partial_{\mu}f(l)l^{\mu}-\Big(\alpha\frac{l^2-p_1^2}{(l+p_1)^2}+\beta\frac{l^2-p_2^2}{(l+p_2)^2}+\alpha+\beta+2-D\Big)\Big)\frac{f(l)}{(l+p_1)^{2\alpha}(l+p_2)^{2\beta}l^2}
\pnt
\end{aligned}
\end{equation}
For $f(l)=1$ we recover the scalar rule of~\cite{Broadhurst:1985vq}
\vspace{0.5cm}
\newcommand{\figwidth}{18}
\newcommand{\figheight}{13}
\begin{equation}
\label{scalar-triangle}
\scriptsize
\settoheight{\eqoff}{$\times$}%
\setlength{\eqoff}{0.5\eqoff}%
\addtolength{\eqoff}{-6.5\unitlength}%
\raisebox{\eqoff}{%
\fmfframe(0,0)(0,0){%
\begin{fmfchar*}(\figwidth,\figheight)
\fmfleft{in}
\fmfright{out}
\fmf{plain,tension=1}{in,v1}
\fmfforce{(0,0.5h)}{in}
\fmfforce{(1.1w,0.5h)}{out}
\fmfforce{(0.5w,0)}{v3}
\fmfforce{(0.5w,h)}{v2}
\fmfforce{(0.03w,0.5h)}{v1}
\fmfforce{(w,0.5h)}{v4}
\fmfforce{(1.5w,0.5h)}{v5}
\fmfforce{(1.7w,0.5h)}{v6}
\fmffixed{(0.5w,0)}{v2,v7}
\fmffixed{(0.5w,0)}{v3,v8}
\fmffreeze
\fmf{plain,label=$\alpha$,l.dist=2,l.side=left}{v1,v2}
\fmf{plain,label=$\beta$,l.dist=2,l.side=right}{v1,v3}
\fmf{plain}{v2,v3}
\fmf{plain}{v2,v7}
\fmf{plain}{v3,v8}
\end{fmfchar*}}}\ \ 
\normalsize
= \Delta(\alpha,\beta)\ \ 
\raisebox{\eqoff}{%
\fmfframe(0,0)(0,0){%
\scriptsize
\begin{fmfchar*}(\figwidth,\figheight)
\fmfleft{in}
\fmfright{out}
\fmf{plain,label=$h(\alpha,,\beta)$,label.side=left,label.dist=2,tension=1}{in,v1}
\fmfforce{(0,0.5h)}{in}
\fmfforce{(1.1w,0.5h)}{out}
\fmfforce{(w,0)}{v3}
\fmfforce{(w,h)}{v2}
\fmfforce{(0.5w,0.5h)}{v1}
\fmfforce{(w,0.5h)}{v4}
\fmfforce{(1.5w,0.5h)}{v5}
\fmfforce{(1.7w,0.5h)}{v6}
\fmffixed{(0.5w,0)}{v2,v7}
\fmffixed{(0.5w,0)}{v3,v8}
\fmffreeze
\fmf{plain}{v1,v2}
\fmf{plain}{v1,v3}
\end{fmfchar*}}}\ \ 
\normalsize
+C(\alpha,\beta)\ \ 
\raisebox{\eqoff}{%
\fmfframe(0,0)(0,0){%
\scriptsize
\begin{fmfchar*}(\figwidth,\figheight)
\fmfleft{in}
\fmfright{out}
\fmf{plain,tension=1}{in,v1}
\fmfforce{(0,0.5h)}{in}
\fmfforce{(1.1w,0.5h)}{out}
\fmfforce{(0.5w,0)}{v3}
\fmfforce{(0.5w,h)}{v2}
\fmfforce{(0.03w,0.5h)}{v1}
\fmfforce{(w,0.5h)}{v4}
\fmfforce{(1.5w,0.5h)}{v5}
\fmfforce{(1.7w,0.5h)}{v6}
\fmffixed{(0.5w,0)}{v2,v7}
\fmffixed{(0.5w,0)}{v3,v8}
\fmffreeze
\fmf{plain,label=$\alpha+1$,l.dist=2,l.side=left}{v1,v2}
\fmf{plain,label=$\beta$,l.dist=2,l.side=right}{v1,v3}
\fmf{plain}{v2,v3}
\fmf{plain}{v3,v8}
\end{fmfchar*}}}\ \ 
\normalsize
+C(\beta,\alpha)\ \ 
\raisebox{\eqoff}{%
\fmfframe(0,0)(0,0){%
\scriptsize
\begin{fmfchar*}(\figwidth,\figheight)
\fmfleft{in}
\fmfright{out}
\fmf{plain,tension=1}{in,v1}
\fmfforce{(0,0.5h)}{in}
\fmfforce{(1.1w,0.5h)}{out}
\fmfforce{(0.5w,0)}{v3}
\fmfforce{(0.5w,h)}{v2}
\fmfforce{(0.03w,0.5h)}{v1}
\fmfforce{(w,0.5h)}{v4}
\fmfforce{(1.5w,0.5h)}{v5}
\fmfforce{(1.7w,0.5h)}{v6}
\fmffixed{(0.5w,0)}{v2,v7}
\fmffixed{(0.5w,0)}{v3,v8}
\fmffreeze
\fmf{plain,label=$\alpha$,l.dist=2,l.side=left}{v1,v2}
\fmf{plain,label=$\beta+1$,l.dist=2,l.side=right}{v1,v3}
\fmf{plain}{v2,v3}
\fmf{plain}{v2,v7}
\end{fmfchar*}}}
\col
\vspace{0.5cm}
\end{equation}
where 
\begin{equation}
\begin{aligned}
\Delta(\alpha,\beta)&=-\frac{\alpha G(\alpha+1,\beta)+\beta G(\alpha,\beta+1)}{\alpha+\beta+2-D}\col\\
C(\alpha,\beta)&=\frac{\alpha}{\alpha+\beta+2-D}\col\\
h(\alpha,\beta)&=\alpha+\beta+1-D/2\col\\
G(\alpha,\beta)&=\frac{1}{(4\pi)^{D/2}}\int\frac{\mathrm{d}^D l}{l^{2\alpha}(l+p)^{2\beta}}\Big\vert_{p^2=1}=\frac{\Gamma(\alpha+\beta-D/2)\Gamma(D/2-\alpha)\Gamma(D/2-\beta)}{\Gamma(\alpha)\Gamma(\beta)\Gamma(D-\alpha-\beta)}
\pnt\end{aligned}
\end{equation}

\vspace{1cm}
If we take $f(l)=l^{\nu}$ we obtain the first generalized rule
\begin{equation}
\label{der1-triangle}
\begin{aligned}
\scriptsize
\settoheight{\eqoff}{$\times$}%
\setlength{\eqoff}{0.5\eqoff}%
\addtolength{\eqoff}{-6.5\unitlength}%
\settoheight{\eqofftwo}{$\times$}%
\setlength{\eqofftwo}{0.5\eqofftwo}%
\addtolength{\eqofftwo}{-7.5\unitlength}%
\raisebox{\eqoff}{%
\begin{fmfchar*}(\figwidth,\figheight)
\fmfleft{in}
\fmfright{out}
\fmf{plain,tension=1}{in,v1}
\fmfforce{(0,0.5h)}{in}
\fmfforce{(1.1w,0.5h)}{out}
\fmfforce{(0.5w,0)}{v3}
\fmfforce{(0.5w,h)}{v2}
\fmfforce{(0.03w,0.5h)}{v1}
\fmfforce{(w,0.5h)}{v4}
\fmfforce{(1.5w,0.5h)}{v5}
\fmfforce{(1.7w,0.5h)}{v6}
\fmffixed{(0.5w,0)}{v2,v7}
\fmffixed{(0.5w,0)}{v3,v8}
\fmffreeze
\fmf{plain,label=$\alpha$,l.dist=2,l.side=left}{v1,v2}
\fmf{plain,label=$\beta$,l.dist=2,l.side=right}{v1,v3}
\fmf{derplain}{v2,v3}
\fmf{plain}{v2,v7}
\fmf{plain}{v3,v8}
\end{fmfchar*}}\ \  
&= \Delta_-(\alpha,\beta)\ \ 
\settoheight{\eqoff}{$\times$}%
\setlength{\eqoff}{0.5\eqoff}%
\addtolength{\eqoff}{-6.5\unitlength}%
\raisebox{\eqoff}{%
\scriptsize
\begin{fmfchar*}(\figwidth,\figheight)
\fmfleft{in}
\fmfright{out}
\fmf{plain,label=$h(\alpha,,\beta)$,label.side=left,label.dist=2,tension=1}{in,v1}
\fmfforce{(0,0.5h)}{in}
\fmfforce{(1.1w,0.5h)}{out}
\fmfforce{(w,0)}{v3}
\fmfforce{(w,h)}{v2}
\fmfforce{(0.5w,0.5h)}{v1}
\fmfforce{(w,0.5h)}{v4}
\fmfforce{(1.5w,0.5h)}{v5}
\fmfforce{(1.7w,0.5h)}{v6}
\fmffixed{(0.5w,0)}{v2,v7}
\fmffixed{(0.5w,0)}{v3,v8}
\fmffreeze
\fmf{plain}{v1,v2}
\fmf{derplain}{v3,v1}
\end{fmfchar*}}\ \ 
\normalsize
- \ \Delta_+(\alpha,\beta)\ \ 
\raisebox{\eqoff}{%
\scriptsize
\begin{fmfchar*}(\figwidth,\figheight)
\fmfleft{in}
\fmfright{out}
\fmf{plain,label=$h(\alpha,,\beta)$,label.side=left,label.dist=2,tension=1}{in,v1}
\fmfforce{(0,0.5h)}{in}
\fmfforce{(1.1w,0.5h)}{out}
\fmfforce{(w,0)}{v3}
\fmfforce{(w,h)}{v2}
\fmfforce{(0.5w,0.5h)}{v1}
\fmfforce{(w,0.5h)}{v4}
\fmfforce{(1.5w,0.5h)}{v5}
\fmfforce{(1.7w,0.5h)}{v6}
\fmffixed{(0.5w,0)}{v2,v7}
\fmffixed{(0.5w,0)}{v3,v8}
\fmffreeze
\fmf{derplain}{v1,v2}
\fmf{plain}{v3,v1}
\end{fmfchar*}}\ \ 
\\
\\
\normalsize
&\qquad+\tilde{C}(\alpha,\beta)\ \ 
\settoheight{\eqoff}{$\times$}%
\setlength{\eqoff}{0.5\eqoff}%
\addtolength{\eqoff}{-6.5\unitlength}%
\raisebox{\eqoff}{%
\scriptsize
\begin{fmfchar*}(\figwidth,\figheight)
\fmfleft{in}
\fmfright{out}
\fmf{plain,tension=1}{in,v1}
\fmfforce{(0,0.5h)}{in}
\fmfforce{(1.1w,0.5h)}{out}
\fmfforce{(0.5w,0)}{v3}
\fmfforce{(0.5w,h)}{v2}
\fmfforce{(0.03w,0.5h)}{v1}
\fmfforce{(w,0.5h)}{v4}
\fmfforce{(1.5w,0.5h)}{v5}
\fmfforce{(1.7w,0.5h)}{v6}
\fmffixed{(0.5w,0)}{v2,v7}
\fmffixed{(0.5w,0)}{v3,v8}
\fmffreeze
\fmf{plain,label=$\alpha+1$,l.dist=2,l.side=left}{v1,v2}
\fmf{plain,label=$\beta$,l.dist=2,l.side=right}{v1,v3}
\fmf{derplain}{v2,v3}
\fmf{plain}{v3,v8}
\end{fmfchar*}}\ \ 
\normalsize
+\tilde{C}(\beta,\alpha)\ \ 
\raisebox{\eqoff}{%
\scriptsize
\begin{fmfchar*}(\figwidth,\figheight)
\fmfleft{in}
\fmfright{out}
\fmf{plain,tension=1}{in,v1}
\fmfforce{(0,0.5h)}{in}
\fmfforce{(1.1w,0.5h)}{out}
\fmfforce{(0.5w,0)}{v3}
\fmfforce{(0.5w,h)}{v2}
\fmfforce{(0.03w,0.5h)}{v1}
\fmfforce{(w,0.5h)}{v4}
\fmfforce{(1.5w,0.5h)}{v5}
\fmfforce{(1.7w,0.5h)}{v6}
\fmffixed{(0.5w,0)}{v2,v7}
\fmffixed{(0.5w,0)}{v3,v8}
\fmffreeze
\fmf{plain,label=$\alpha$,l.dist=2,l.side=left}{v1,v2}
\fmf{plain,label=$\beta+1$,l.dist=2,l.side=right}{v1,v3}
\fmf{derplain}{v2,v3}
\fmf{plain}{v2,v7}
\end{fmfchar*}}
\col
\end{aligned}
\vspace{0.5cm}
\end{equation}
where 
\begin{equation}
\label{rules-def}
\begin{aligned}
\Delta_{\pm}(\alpha,\beta)&=\Delta_1(\alpha,\beta)\pm\tilde{\Delta}(\alpha,\beta)\col\\
\Delta_1(\alpha,\beta)&=\frac{(\alpha-\beta)G(\alpha,\beta)-\alpha G(\alpha+1,\beta-1)+\beta G(\alpha-1,\beta+1)}{2(\alpha+\beta+1-D)}\col\\
\tilde{\Delta}(\alpha,\beta)&=-\frac{\alpha G(\alpha+1,\beta)+\beta G(\alpha,\beta+1)}{2(\alpha+\beta+1-D)}\col\\
\tilde{C}(\alpha,\beta)&=\frac{\alpha}{\alpha+\beta+1-D}
\pnt
\end{aligned}
\end{equation}
Another useful formula can be derived from the previous two:
\begin{equation}
\label{der2-triangle}
\begin{aligned}
\scriptsize
\settoheight{\eqoff}{$\times$}%
\setlength{\eqoff}{0.5\eqoff}%
\addtolength{\eqoff}{-6.5\unitlength}%
\settoheight{\eqofftwo}{$\times$}%
\setlength{\eqofftwo}{0.5\eqofftwo}%
\addtolength{\eqofftwo}{-7.5\unitlength}%
\raisebox{\eqoff}{%
\begin{fmfchar*}(\figwidth,\figheight)
\fmfleft{in}
\fmfright{out}
\fmf{plain,tension=1}{in,v1}
\fmfforce{(0,0.5h)}{in}
\fmfforce{(1.1w,0.5h)}{out}
\fmfforce{(0.5w,0)}{v3}
\fmfforce{(0.5w,h)}{v2}
\fmfforce{(0.03w,0.5h)}{v1}
\fmfforce{(w,0.5h)}{v4}
\fmfforce{(1.5w,0.5h)}{v5}
\fmfforce{(1.7w,0.5h)}{v6}
\fmffixed{(0.5w,0)}{v2,v7}
\fmffixed{(0.5w,0)}{v3,v8}
\fmffreeze
\fmf{derplain,label=$\alpha$,l.dist=2,l.side=left}{v1,v2}
\fmf{plain,label=$\beta$,l.dist=2,l.side=right}{v1,v3}
\fmf{plain}{v2,v3}
\fmf{plain}{v2,v7}
\fmf{plain}{v3,v8}
\end{fmfchar*}}\ \  
&= -\big(\Delta_1(\alpha,\beta)-\frac{1}{2}\Delta(\alpha,\beta)\big)\ \ 
\settoheight{\eqoff}{$\times$}%
\setlength{\eqoff}{0.5\eqoff}%
\addtolength{\eqoff}{-6.5\unitlength}%
\raisebox{\eqoff}{%
\scriptsize
\begin{fmfchar*}(\figwidth,\figheight)
\fmfleft{in}
\fmfright{out}
\fmf{derplain,label=$h(\alpha,,\beta)$,label.side=left,label.dist=2,tension=1}{in,v1}
\fmfforce{(0,0.5h)}{in}
\fmfforce{(1.1w,0.5h)}{out}
\fmfforce{(w,0)}{v3}
\fmfforce{(w,h)}{v2}
\fmfforce{(0.5w,0.5h)}{v1}
\fmfforce{(w,0.5h)}{v4}
\fmfforce{(1.5w,0.5h)}{v5}
\fmfforce{(1.7w,0.5h)}{v6}
\fmffixed{(0.5w,0)}{v2,v7}
\fmffixed{(0.5w,0)}{v3,v8}
\fmffreeze
\fmf{plain}{v1,v2}
\fmf{plain}{v3,v1}
\end{fmfchar*}}
\\
\normalsize
&\qquad- \ \frac{\Delta(\alpha,\beta)}{2(\alpha+\beta+1-D)}\Big(\ \ 
\settoheight{\eqoff}{$\times$}%
\setlength{\eqoff}{0.5\eqoff}%
\addtolength{\eqoff}{-6.5\unitlength}%
\raisebox{\eqoff}{%
\scriptsize
\begin{fmfchar*}(\figwidth,\figheight)
\fmfleft{in}
\fmfright{out}
\fmf{plain,label=$h(\alpha,,\beta)$,label.side=left,label.dist=2,tension=1}{in,v1}
\fmfforce{(0,0.5h)}{in}
\fmfforce{(1.1w,0.5h)}{out}
\fmfforce{(w,0)}{v3}
\fmfforce{(w,h)}{v2}
\fmfforce{(0.5w,0.5h)}{v1}
\fmfforce{(w,0.5h)}{v4}
\fmfforce{(1.5w,0.5h)}{v5}
\fmfforce{(1.7w,0.5h)}{v6}
\fmffixed{(0.5w,0)}{v2,v7}
\fmffixed{(0.5w,0)}{v3,v8}
\fmffreeze
\fmf{derplain}{v1,v2}
\fmf{plain}{v3,v1}
\end{fmfchar*}}\ \ 
\normalsize
+ \ \  
\settoheight{\eqoff}{$\times$}%
\setlength{\eqoff}{0.5\eqoff}%
\addtolength{\eqoff}{-6.5\unitlength}%
\raisebox{\eqoff}{%
\scriptsize
\begin{fmfchar*}(\figwidth,\figheight)
\fmfleft{in}
\fmfright{out}
\fmf{plain,label=$h(\alpha,,\beta)$,label.side=left,label.dist=2,tension=1}{in,v1}
\fmfforce{(0,0.5h)}{in}
\fmfforce{(1.1w,0.5h)}{out}
\fmfforce{(w,0)}{v3}
\fmfforce{(w,h)}{v2}
\fmfforce{(0.5w,0.5h)}{v1}
\fmfforce{(w,0.5h)}{v4}
\fmfforce{(1.5w,0.5h)}{v5}
\fmfforce{(1.7w,0.5h)}{v6}
\fmffixed{(0.5w,0)}{v2,v7}
\fmffixed{(0.5w,0)}{v3,v8}
\fmffreeze
\fmf{plain}{v1,v2}
\fmf{derplain}{v3,v1}
\end{fmfchar*}}
\ \ \Big)
\\
\\
\normalsize
&\qquad+\big(\tilde{C}(\alpha,\beta)-C(\alpha,\beta)\big)\ \ 
\settoheight{\eqoff}{$\times$}%
\setlength{\eqoff}{0.5\eqoff}%
\addtolength{\eqoff}{-6.5\unitlength}%
\raisebox{\eqoff}{%
\scriptsize
\begin{fmfchar*}(\figwidth,\figheight)
\fmfleft{in}
\fmfright{out}
\fmf{plain,tension=1}{in,v1}
\fmfforce{(0,0.5h)}{in}
\fmfforce{(1.1w,0.5h)}{out}
\fmfforce{(0.5w,0)}{v3}
\fmfforce{(0.5w,h)}{v2}
\fmfforce{(0.03w,0.5h)}{v1}
\fmfforce{(w,0.5h)}{v4}
\fmfforce{(1.5w,0.5h)}{v5}
\fmfforce{(1.7w,0.5h)}{v6}
\fmffixed{(0.5w,0)}{v2,v7}
\fmffixed{(0.5w,0)}{v3,v8}
\fmffreeze
\fmf{plain,label=$\alpha+1$,l.dist=2,l.side=left}{v1,v2}
\fmf{plain,label=$\beta$,l.dist=2,l.side=right}{v1,v3}
\fmf{derplain}{v2,v3}
\fmf{plain}{v3,v8}
\end{fmfchar*}}\ \ 
\normalsize
+\big(\tilde{C}(\beta,\alpha)-C(\beta,\alpha)\big)\ \ 
\raisebox{\eqoff}{%
\scriptsize
\begin{fmfchar*}(\figwidth,\figheight)
\fmfleft{in}
\fmfright{out}
\fmf{plain,tension=1}{in,v1}
\fmfforce{(0,0.5h)}{in}
\fmfforce{(1.1w,0.5h)}{out}
\fmfforce{(0.5w,0)}{v3}
\fmfforce{(0.5w,h)}{v2}
\fmfforce{(0.03w,0.5h)}{v1}
\fmfforce{(w,0.5h)}{v4}
\fmfforce{(1.5w,0.5h)}{v5}
\fmfforce{(1.7w,0.5h)}{v6}
\fmffixed{(0.5w,0)}{v2,v7}
\fmffixed{(0.5w,0)}{v3,v8}
\fmffreeze
\fmf{plain,label=$\alpha$,l.dist=2,l.side=left}{v1,v2}
\fmf{plain,label=$\beta+1$,l.dist=2,l.side=right}{v1,v3}
\fmf{derplain}{v2,v3}
\fmf{plain}{v2,v7}
\end{fmfchar*}}
\\
\normalsize
&\qquad+C(\alpha,\beta)\ \ 
\settoheight{\eqoff}{$\times$}%
\setlength{\eqoff}{0.5\eqoff}%
\addtolength{\eqoff}{-6.5\unitlength}%
\raisebox{\eqoff}{%
\scriptsize
\begin{fmfchar*}(\figwidth,\figheight)
\fmfleft{in}
\fmfright{out}
\fmf{plain,tension=1}{in,v1}
\fmfforce{(0,0.5h)}{in}
\fmfforce{(1.1w,0.5h)}{out}
\fmfforce{(0.5w,0)}{v3}
\fmfforce{(0.5w,h)}{v2}
\fmfforce{(0.03w,0.5h)}{v1}
\fmfforce{(w,0.5h)}{v4}
\fmfforce{(1.5w,0.5h)}{v5}
\fmfforce{(1.7w,0.5h)}{v6}
\fmffixed{(0.5w,0)}{v2,v7}
\fmffixed{(0.5w,0)}{v3,v8}
\fmffreeze
\fmf{derplain,label=$\alpha+1$,l.dist=2,l.side=left}{v1,v2}
\fmf{plain,label=$\beta$,l.dist=2,l.side=right}{v1,v3}
\fmf{plain}{v2,v3}
\fmf{plain}{v3,v8}
\end{fmfchar*}}\ \ 
\normalsize
+C(\beta,\alpha)\ \ 
\raisebox{\eqoff}{%
\scriptsize
\begin{fmfchar*}(\figwidth,\figheight)
\fmfleft{in}
\fmfright{out}
\fmf{plain,tension=1}{in,v1}
\fmfforce{(0,0.5h)}{in}
\fmfforce{(1.1w,0.5h)}{out}
\fmfforce{(0.5w,0)}{v3}
\fmfforce{(0.5w,h)}{v2}
\fmfforce{(0.03w,0.5h)}{v1}
\fmfforce{(w,0.5h)}{v4}
\fmfforce{(1.5w,0.5h)}{v5}
\fmfforce{(1.7w,0.5h)}{v6}
\fmffixed{(0.5w,0)}{v2,v7}
\fmffixed{(0.5w,0)}{v3,v8}
\fmffreeze
\fmf{derplain,label=$\alpha$,l.dist=2,l.side=left}{v1,v2}
\fmf{plain,label=$\beta+1$,l.dist=2,l.side=right}{v1,v3}
\fmf{plain}{v2,v3}
\fmf{plain}{v2,v7}
\end{fmfchar*}}
\pnt
\end{aligned}
\vspace{0.5cm}
\end{equation}

\subsection{Example of recurrence relation}
\label{app:example-rec}

As an example of recurrence relation obtained from the triangle rules, let us consider the computation of $I_L^{(1)}$. We can write
\begin{equation}
I_L^{(1)}=K_L^{(1)}(1,\ldots,1)
\col
\end{equation}
where $K_L^{(1)}(\a_1,\ldots,\a_L)$ is the integral with the same topology and numerator as $I_L^{(1)}$, but weights $\a_1,\ldots,\a_L$ on the radial lines. Using~\eqref{scalar-triangle} we find
\begin{equation}
\begin{aligned}
{}&K_L^{(1)}(\a_1,\ldots,\a_L)=\Delta(\a_{L-2},\a_{L-1})K_{L-1}^{(1)}(\a_1,\ldots,\a_{L-3},h(a_{L-2},\a_{L-1}),\a_L)\\
&\qquad\quad+C(\a_{L-2},\a_{L-1})G(\a_{L-3},\a_{L-2}+1)K_{L-1}^{(1)}(\a_1,\ldots,g(\a_{L-3},\a_{L-2}+1),\a_{L-1},\a_L)\\
&\qquad\quad+C(\a_{L-1},\a_{L-2})G(\a_{L-1}+1,\a_{L})K_{L-1}^{(1)}(\a_1,\ldots,\a_{L-2},g(\a_{L-1}+1,\a_{L}))
\col
\end{aligned}
\end{equation}
where $g(\alpha,\beta)=\alpha+\beta-D/2$ and $h$ is defined in eq.~\eqref{rules-def}.\\
We can go on applying eq.~\eqref{scalar-triangle} until we obtain $K_5^{(1)}$ in terms of $K_4^{(1)}$. Then, using the generalized rules, we can compute $K_4^{(1)}$ explicitly.\\
One can deal with integrals $I_L^{(2)}$ and $I_L^{(3)}$ in the same way, using eq.~\eqref{scalar-triangle} down to five and six loops respectively. For the general case $I_L^{(j)}$ with $j\geq4$, eq.~\eqref{scalar-triangle} must be applied on both sides of the integrals to reduce them to seven loops.

\subsection{GPXT in $p$-space}
\label{app:GPXT}

We are interested in the pole part of the integrals $I_L^{(j)}$ and $P_L$ 
given in Fig.~\ref{integrals}. 
Thereby, the relation \eqref{ILrel} implies that the 
independent non-vanishing integrals are represented by
$1\le j\le[\frac{L}{2}]$, where $[x]$ denotes the integer part of $x$

Instead of solving the integrals in $x$-space, we work directly in 
$p$-space. This has significant advantages: one does not have to shift the 
derivatives to the root vertex. Therefore, one does not encounter the
problem to compute several integrals with subdivergences which combine to 
the required integral. Furthermore, the number of integration points is 
reduced by one. The calculation can therefore be pushed to higher loop order.
We first write the required integral as the following linear combination
\begin{equation}
I_L^{(j)}=J_L^{(j)}-J_L^{(j+1)}\col
\end{equation}
where
\begin{equation}
\begin{aligned}
\unitlength=0.75mm
J_L^{(k)}=
\settoheight{\eqoff}{$\times$}%
\setlength{\eqoff}{0.5\eqoff}%
\addtolength{\eqoff}{-15\unitlength}%
\raisebox{\eqoff}{%
\fmfframe(0,0)(10,0){%
\begin{fmfchar*}(30,30)
  \fmfleft{in}
  \fmfright{out}
  \fmf{plain,tension=1}{in,vi}
  \fmf{phantom,tension=1}{out,v5}
\fmfpoly{phantom}{vc,v9,v8,v7,v6,v5,v4,v3,v2,v1,vb,va}
\fmffixed{(0.9w,0)}{va,v5}
\fmffixed{(0.3w,0)}{vi,v0}
\fmffixed{(0,whatever)}{v0,v1}
\fmf{dashes}{v1,v2}
\fmf{dashes}{v2,v3}
\fmf{derplain}{v3,v4}
\fmf{plain}{v4,v5}
\fmf{dashes}{v5,v6}
\fmf{dashes}{v6,v7}
\fmf{plain}{v7,v8}
\fmf{plain}{v8,v9}
\fmf{derplain}{vi,v1}
\fmf{plain}{vi,v9}
\fmf{plain}{v0,v1}
\fmf{dashes}{v0,v2}
\fmf{plain}{v0,v3}
\fmf{plain}{v0,v4}
\fmf{plain}{v0,v5}
\fmf{dashes}{v0,v6}
\fmf{plain}{v0,v7}
\fmf{plain}{v0,v8}
\fmf{plain}{v0,v9}
\fmffreeze
\fmfiv{l=\scriptsize $x_1$,l.d=2}{vloc(__v1)}
\fmfiv{l=\scriptsize $x_{k-1}$,l.d=2}{vloc(__v3)}
\fmfiv{l=\scriptsize $x_k$,l.d=2}{vloc(__v4)}
\fmfiv{l=\scriptsize $x_{k+1}$,l.d=2}{vloc(__v5)}
\fmfiv{l=\scriptsize $x_L$,l.d=2}{vloc(__v9)}
\end{fmfchar*}}}
=
\frac{1}{(2\pi)^{LD}}\int\frac{\de^Dp_1\dots\de^Dp_L\,p_1\cdot p_k}
{p_1^4p_2^2\dots p_L^2(p_1-p_2)^2\dots (p_{L-1}-p_L)^2(p_L-p_1)^2}
\pnt
\end{aligned}
\end{equation}
The $p$-space graphs of the integrals $J_L^{(k)}$ which 
we need for $1\le k\le K$, $K=[\frac{L}{2}]+1$
 hence read
\begin{equation}
\begin{aligned}
\unitlength=0.75mm
J_L^{(1)}=P_L=
\settoheight{\eqoff}{$\times$}%
\setlength{\eqoff}{0.5\eqoff}%
\addtolength{\eqoff}{-15\unitlength}%
\raisebox{\eqoff}{%
\fmfframe(4,0)(0,0){%
\begin{fmfchar*}(30,30)
  \fmfleft{in}
  \fmfright{out}
  \fmf{phantom,tension=1}{in,v1}
  \fmf{phantom,tension=1}{in,v10}
  \fmf{phantom,tension=1}{out,v4}
  \fmf{phantom,tension=1}{out,v5}
\fmfpoly{phantom}{v9,v8,v7,v6,v5,v4,v3,v2,v1}
\fmffixed{(0.05w,0)}{in,v1}
\fmffixed{(0.05w,whatever)}{v5,out}
\fmffixed{(0,whatever)}{v5,v6}
\fmf{plain}{v1,v2}
\fmf{dashes}{v2,v3}
\fmf{plain}{v3,v4}
\fmf{plain}{v4,v5}
\fmf{plain}{v5,v6}
\fmf{dashes}{v6,v7}
\fmf{plain}{v7,v8}
\fmf{plain}{v8,v9}
\fmf{plain}{v9,v1}
\fmf{plain}{v0,v1}
\fmf{dashes}{v0,v2}
\fmf{dashes}{v0,v3}
\fmf{plain}{v0,v4}
\fmf{plain}{v0,v5}
\fmf{dashes}{v0,v6}
\fmf{dashes}{v0,v7}
\fmf{plain}{v0,v8}
\fmf{plain}{v0,v9}
\fmffreeze
\fmfiv{l=\scriptsize $p_1$,l.d=2}{vloc(__v1)}
\fmfiv{l=\scriptsize $p_L$,l.d=2}{vloc(__v9)}
\end{fmfchar*}}}
\col\qquad
J_L^{(k)}=
\settoheight{\eqoff}{$\times$}%
\setlength{\eqoff}{0.5\eqoff}%
\addtolength{\eqoff}{-15\unitlength}%
\raisebox{\eqoff}{%
\fmfframe(4,0)(6,0){%
\begin{fmfchar*}(30,30)
  \fmfleft{in}
  \fmfright{out}
  \fmf{phantom,tension=1}{in,v1}
  \fmf{phantom,tension=1}{in,v10}
  \fmf{phantom,tension=1}{out,v4}
  \fmf{phantom,tension=1}{out,v5}
\fmfpoly{phantom}{v9,v8,v7,v6,v5,v4,v3,v2,v1}
\fmffixed{(0.05w,0)}{in,v1}
\fmffixed{(0.05w,whatever)}{v5,out}
\fmffixed{(0,whatever)}{v5,v6}
\fmf{plain}{v1,v2}
\fmf{dashes}{v2,v3}
\fmf{plain}{v3,v4}
\fmf{plain}{v4,v5}
\fmf{plain}{v5,v6}
\fmf{dashes}{v6,v7}
\fmf{plain}{v7,v8}
\fmf{plain}{v8,v9}
\fmf{plain}{v9,v1}
\fmf{derplain,label=$\scriptscriptstyle 2$,l.side=right,l.dist=3}{v0,v1}
\fmf{dashes}{v0,v2}
\fmf{dashes}{v0,v3}
\fmf{derplain}{v0,v4}
\fmf{plain}{v0,v5}
\fmf{dashes}{v0,v6}
\fmf{dashes}{v0,v7}
\fmf{plain}{v0,v8}
\fmf{plain}{v0,v9}
\fmffreeze
\fmfiv{l=\scriptsize $p_1$,l.d=2}{vloc(__v1)}
\fmfiv{l=\scriptsize $p_{k-1}$,l.d=2}{vloc(__v3)}
\fmfiv{l=\scriptsize $p_k$,l.d=2}{vloc(__v4)}
\fmfiv{l=\scriptsize $p_{k+1}$,l.d=2}{vloc(__v5)}
\fmfiv{l=\scriptsize $p_L$,l.d=2}{vloc(__v9)}
\end{fmfchar*}}}
\pnt
\end{aligned}
\end{equation}
Where the number at a line denotes the weight of the propagator. 
We apply GPXT \cite{Chetyrkin:1980pr} to the above integrals in $p$-space. The
propagators are first expanded in terms of Gegenbauer polynomials $C_i^{(1)}(x)$ (which in this case are the Chebyshev polynomials of the second kind). 
Then, the angular integrals are performed. This yields
\begin{equation}
\begin{aligned}
\label{Jint}
J_L^{(1)}&=\frac{1}{(2\pi)^{LD}}\frac{\Omega_{D-1}^L}{2^L}
\sum_{i=0}^\infty\Big(\frac{1}{i+1}\Big)^{L-1}
R_\lambda^{(1)}(i)\col\\
J_L^{(k)}&=\frac{1}{(2\pi)^{LD}}\frac{\Omega_{D-1}^L}{2^L}\frac{1}{2}
\sum_{i=0}^\infty\sum_{\substack{i=|j-1|\\i\neq j}}^{j+1}
\Big(\frac{1}{i+1}\Big)^{L-k+1}\Big(\frac{1}{j+1}\Big)^{k-2}D_1(j,1,i)C_i^{(1)}(1)
R_\lambda^{(k)}(i,j)\col
\end{aligned}
\end{equation}
where $\Omega_{D-1}$ is the volume of the $(D-1)$-dimensional unit sphere, 
$D_1(j,1,i)$ are the Clebsch-Gordan coefficients of the Gegenbauer polynomials, 
and the radial integrals are given by
\begin{equation}
\begin{aligned}
\label{Rint}
R_\lambda^{(1)}(i)&=
\int_R^\infty\frac{\de r_1\dots\de r_L\,
(r_1\dots r_L)^{\lambda-1}}
{\max_{r_1r_2}\dots\max_{r_{L-1}r_L}\max_{r_Lr_1}}
\Big(\frac{\min_{r_1r_2}}{\max_{r_1r_2}}\dots\frac{\min_{r_{L-1}r_L}}{\max_{r_{L-1}r_L}}\frac{\min_{r_Lr_1}}{\max_{r_Lr_1}}\Big)^{\frac{i}{2}}\col\\
R_\lambda^{(k)}(i,j)&=
\int_R^\infty\frac{\de r_1\dots\de r_L\,r_1^{\lambda-\frac{3}{2}}
r_k^{\lambda-\frac{1}{2}} (r_2\dots r_{k-1}r_{k+1}\dots r_L)^{\lambda-1}}
{\max_{r_1r_2}\dots\max_{r_{L-1}r_L}\max_{r_Lr_1}}\\
&\phantom{{}={}\int_P^\infty}
\Big(\frac{\min_{r_1r_2}}{\max_{r_1r_2}}\dots\frac{\min_{r_{k-1}r_k}}{\max_{r_{k-1}r_k}}\Big)^{\frac{j}{2}}
\Big(\frac{\min_{r_kr_{k+1}}}{\max_{r_kr_{k+1}}}\dots\frac{\min_{r_{L-1}r_L}}{\max_{r_{L-1}r_L}}\frac{\min_{r_Lr_1}}{\max_{r_Lr_1}}\Big)^{\frac{i}{2}}
\pnt
\end{aligned}
\end{equation}
we have introduced a regulator $R$ as a lower bound for the momentum integration. This cuts out the infrared regime of the integrals and hence does not affect
the pole part we are interested in\footnote{In this case, where no infrared divergences are present, one can safely set $R=0$ as long as one keeps factors $R^{\lambda-1}=R^{-\varepsilon}\to1$}. While it is simple to directly evaluate the 
above integrals for small $L$, it becomes very tedious at larger $L$.
We found that it is advantageous to set up a recurrence relation for the
radial integrals.

The starting point is the integral
\begin{equation}
\begin{aligned}
&I_1(a_1,b_1,b_2,c_1,c_2;r_0,r_2)
=\int_R^\infty\de r_1\,r_1^{a_1}
{\textstyle\max}_{r_0r_1}^{b_1}{\textstyle\max}_{r_1r_2}^{b_2}
{\textstyle\min}_{r_0r_1}^{c_1}{\textstyle\min}_{r_1r_2}^{c_2}\Big|_{r_0\ge r_2}\\
&=-\frac{R^{a_1+c_1+c_2+1}r_0^{b_1}r_2^{b_2}}{a_1+c_1+c_2+1}
+\frac{r_0^{b_1}r_2^{c_2}}{a_1+b_2+c_1+1}
\Big(\frac{(b_1-c_1)r_0^{a_1+b_2+c_1+1}}{a_1+b_1+b_2+1}
+\frac{(b_2-c_2)r_2^{a_1+b_2+c_1+1}}{a_1+c_1+c_2+1}\Big)
\col
\end{aligned}
\end{equation}
where we assume that the constants $a_1$, $b_1$, $b_2$, $c_1$, $c_2$ are such that the upper boundary does not contribute.
Longer chains of integration are defined as 
\begin{equation}
\begin{aligned}
&I_n(a_1,\dots,a_n,b_1,\dots,b_{n+1},c_1,\dots,c_{n+1};r_0,r_{n+1})\\
&\qquad\qquad
=\int_R^\infty\de r_1\dots\de r_n\,r_1^{a_1}\dots r_n^{a_n}
{\textstyle\max}_{r_0r_1}^{b_1}{\textstyle\max}_{r_1r_2}^{b_2}\dots {\textstyle\max}_{r_nr_{n+1}}^{b_{n+1}}\\
&\qquad\qquad
\phantom{{}={}\int_R^\infty\de r_1\dots\de r_n\,r_1^{a_1}\dots r_n^{a_n}}
{\textstyle\min}_{r_0r_1}^{c_1}{\textstyle\min}_{r_1r_2}^{c_2}\dots{\textstyle\min}_{r_nr_{n+1}}^{c_{n+1}}\Big|_{r_0\ge r_{n+1}}\pnt
\end{aligned}
\end{equation}
It is important to remark that for the 
remaining domain $r_0<r_{n+1}$ the result is obtained by simply reverting the 
order $(a_1,\dots,a_n)\to(a_n,\dots,a_1)$, 
$(b_1,\dots,b_{n+1})\to(b_{n+1},\dots,b_1)$,
$(c_1,\dots,c_{n+1})\to(c_{n+1},\dots,c_1)$
and by also exchanging 
$r_0\leftrightarrow r_{n+1}$. 

We can then obtain a recurrence relation for the $(n+1)$-fold integral for the 
regime $r_0\ge r_{n+2}$ by explicitly evaluating the integration of
$r_{n+1}$ in 
\begin{equation*}
I_{n+1}(a_1,\dots,a_{n+1},b_1,\dots,b_{n+2},c_1,\dots,c_{n+2};r_0,r_{n+2})
\pnt
\end{equation*}
If we assume again that
there are no contributions from the integrals boundary at infinity, guaranteed
by the values of the constants $a_i$, $b_i$, $c_i$, and that the integrands are
always polynomials, we find the following rule
\begin{equation}
\begin{aligned}
{}&I_{n+1}(a_1,\dots,a_{n+1},b_1,\dots,b_{n+2},c_1,\dots,c_{n+2};r_0,r_{n+2})\\
&=
\big[(r_{n+2}^{b_{n+2}}r_{n+1}^{a_{n+1}+c_{n+2}}-r_{n+1}^{a_{n+1}+b_{n+2}}r_{n+2}^{c_{n+2}})\\
&\phantom{{}={}\big[}
I_n(a_1,\dots,a_n,b_1,\dots,b_{n+1},c_1,\dots,c_{n+1};r_0,r_{n+1})\big]_{r_{n+1}^\alpha\to\frac{1}{\alpha+1}r_{n+2}^{\alpha+1}}\\
&\phantom{{}={}}
-\big[r_{n+1}^{a_{n+1}+c_{n+2}}r_{n+2}^{b_{n+2}}I_n(a_1,\dots,a_n,b_1,\dots,b_{n+1},c_1,\dots,c_{n+1};r_0,r_{n+1})\big]_{r_{n+1}^\alpha\to\frac{1}{\alpha+1}R^{\alpha+1}}\\
&\phantom{{}={}}
+\big[r_{n+1}^{a_{n+1}+b_{n+2}}r_{n+2}^{c_{n+2}}
(I_n(a_1,\dots,a_n,b_1,\dots,b_{n+1},c_1,\dots,c_{n+1};r_0,r_{n+1})\\
&\phantom{{}={}+\big[[r_{n+1}^{a_{n+1}+b_{n+2}}r_{n+2}^{c_{n+2}}(}
-I_n(a_n,\dots,a_1,b_{n+1},\dots,b_1,c_{n+1},\dots,c_1;r_{n+1},r_0))\big]_{r_{n+1}^\alpha\to\frac{1}{\alpha+1}r_0^{\alpha+1}}
\col
\end{aligned} 
\end{equation}
where the replacement means that one first collects all factors of $r_{n+1}$
within each term of the corresponding expression in brackets, and then replaces the appearing factor $r_{n+1}$ with exponent $\alpha$ as indicated. This
mimics the integrations and can be easily evaluated with a computer.
To obtain the final closed chain of integrations, we first 
abbreviate the chain with equal weights as
\begin{equation}
I_n(a,b,c;r_0,r_{n+1})= I_n(a,\dots,a,b,\dots,b,c,\dots,c;r_0,r_{n+1})
\pnt
\end{equation}
Two open chains are then fused by identifying the respective first and second coordinate argument and then integrating over both arguments with appropriate 
additional power factors of these arguments.
Using again the replacement rule to mimic the integrations, the
combinations we need for the radial integrals are then obtained from the 
open chains as follows
\begin{equation}
\begin{aligned}
{}&I_L(a_1,\dots,a_{K},b_1,\dots,b_{K-1},c_1,\dots,c_{K-1},a,b,c)\\
&=\Big[
\big[r_1^{a_1}r_{K}^{a_{K}}
(I_{K-2}(a_2,\dots,a_{K-1},b_1,\dots,b_{K-1},c_1,\dots,c_{K-1};r_1,r_{K})
I_{L-K}(a,b,c;r_1,r_{K})\\
&\phantom{{}={}\Big[}
-I_{K-2}(a_{K-1},\dots,a_2,b_{K-1},\dots,b_1,c_{K-1},\dots,c_1;r_{K},r_1)
I_{L-K}(a,b,c;r_{K},r_1))\big]_{r_1^\alpha\to\frac{1}{\alpha+1}r_{K}^{\alpha+1}}\\
&\phantom{{}={}\Big[}
+
\big[r_1^{a_1}r_{K}^{a_{K}}
I_{K-2}(a_{K-1},\dots,a_2,b_{K-1},\dots,b_1,c_{K-1},\dots,c_1;r_{K},r_1)\\
&\phantom{{}={}\Big[+
\big[}
I_{L-K}(a,b,c;r_{K},r_1)\big]_{r_1^\alpha\to\frac{1}{\alpha+1}R^{\alpha+1}}\Big]_{r_K^\alpha\to\frac{1}{\alpha+1}R^{\alpha+1}}
\pnt
\end{aligned} 
\end{equation}
The required radial integrals $R_\lambda^{(1)}(i)$ and $R_\lambda^{(k)}(i,j)$
in \eqref{Rint}
are then directly given by the above expression with respectively chosen 
constants. 
For  $R_\lambda^{(1)}(i)$ the constants become
\begin{equation}
a_1=\dots=a_K=a=\lambda-1\col\quad 
b_1=\dots=b_{K-1}=b=-1-\frac{i}{2}\col\quad 
c_1=\dots=c_{K-1}=c=\frac{i}{2}
\col
\end{equation}
while for $R_\lambda^{(k)}(i,j)$ we take the above values except for the 
the following constants
\begin{equation}
a_1=\lambda-\frac{3}{2}\col\quad 
a_k=\lambda-\frac{1}{2}\col\quad
b_1=\dots=b_{k-1}=-1-\frac{j}{2}\col\quad
c_1=\dots=c_{k-1}=\frac{j}{2}
\pnt
\end{equation}
With this procedure, we could find analytic expressions for the integrals up 
to $L=11$ which are listed below.

\subsection{Integrals up to $L=11$}
\label{app:results}
Here we show the explicit results for the integrals $I_L^{(j)}$ which are relevant for the computation of anomalous dimensions up to $L=11$. 
\vspace{0.5cm}
\begin{equation*}
\footnotesize
\begin{aligned}
I_4^{(1)}&=\frac{1}{(4\pi)^8}\frac{1}{\varepsilon}\Big[\ \frac{1}{2}\,\zeta(3)+\frac{5}{2}\,\zeta(5)\ \Big]\col\\
I_5^{(1)}&=\frac{1}{(4\pi)^{10}}\frac{1}{\varepsilon}\Big[\ 2\,\zeta(5)+7\,\zeta(7)\ \Big]\col\\
I_6^{(1)}&=\frac{1}{(4\pi)^{12}}\frac{1}{\varepsilon}\Big[\ \frac{2}{3}\,\zeta(5)+\frac{35}{6}\,\zeta(7)+21\,\zeta(9)\ \Big]\col\\
I_7^{(1)}&=\frac{1}{(4\pi)^{14}}\frac{1}{\varepsilon}\Big[\ 3\,\zeta(7)+18\,\zeta(9)+66\,\zeta(11)\ \Big]\col\\
I_8^{(1)}&=\frac{1}{(4\pi)^{16}}\frac{1}{\varepsilon}\Big[\ \frac{3}{4}\,\zeta(7)+\frac{21}{2}\,\zeta(9)+\frac{231}{4}\,\zeta(11)+\frac{429}{2}\,\zeta(13)\ \Big]\col\\
I_9^{(1)}&=\frac{1}{(4\pi)^{18}}\frac{1}{\varepsilon}\Big[\ 4\,\zeta(9)+\frac{110}{3}\,\zeta(11)+\frac{572}{3}\,\zeta(13)+715\,\zeta(15)\ \Big]\col\\
I_{10}^{(1)}&=\frac{1}{(4\pi)^{20}}\frac{1}{\varepsilon}\Big[\ \frac{4}{5}\,\zeta(9)+\frac{33}{2}\,\zeta(11)+\frac{1287}{10}\,\zeta(13)+\frac{1287}{2}\,\zeta(15)+2431\,\zeta(17)\ \Big]\col\\
I_{11}^{(1)}&=\frac{1}{(4\pi)^{18}}\frac{1}{\varepsilon}\Big[\ 5\,\zeta(11)+65\,\zeta(13)+455\,\zeta(15)+2210\,\zeta(17)\ +8398\,\zeta(19)\ \Big]\col\\
& \\
I_4^{(2)}&=\frac{1}{(4\pi)^8}\frac{1}{\varepsilon}\Big[\ -\frac{3}{2}\,\zeta(3)+\frac{5}{2}\,\zeta(5)\ \Big]\col\\
I_5^{(2)}&=\frac{1}{(4\pi)^{10}}\frac{1}{\varepsilon}\Big[\ -4\,\zeta(5)+7\,\zeta(7)\ \Big]\col\\
I_6^{(2)}&=\frac{1}{(4\pi)^{12}}\frac{1}{\varepsilon}\Big[\ -\frac{10}{3}\,\zeta(5)-\frac{49}{6}\,\zeta(7)+21\,\zeta(9)\ \Big]\col\\
I_7^{(2)}&=\frac{1}{(4\pi)^{14}}\frac{1}{\varepsilon}\Big[\ -12\,\zeta(7)-24\,\zeta(9)+66\,\zeta(11)\ \Big]\col\\
I_8^{(2)}&=\frac{1}{(4\pi)^{16}}\frac{1}{\varepsilon}\Big[\ -\frac{21}{4}\,\zeta(7)-\frac{75}{2}\,\zeta(9)-\frac{297}{4}\,\zeta(11)+\frac{429}{2}\,\zeta(13)\ \Big]\col\\
I_9^{(2)}&=\frac{1}{(4\pi)^{18}}\frac{1}{\varepsilon}\Big[\ -24\,\zeta(9)-\frac{385}{3}\,\zeta(11)-\frac{715}{3}\,\zeta(13)+715\,\zeta(15)\ \Big]\col\\
I_{10}^{(2)}&=\frac{1}{(4\pi)^{20}}\frac{1}{\varepsilon}\Big[\ -\frac{36}{5}\,\zeta(9)-\frac{187}{2}\,\zeta(11)-\frac{4433}{10}\,\zeta(13)-\frac{1573}{2}\,\zeta(15)+2431\,\zeta(17)\ \Big]\col\\
I_{11}^{(2)}&=\frac{1}{(4\pi)^{18}}\frac{1}{\varepsilon}\Big[\ -40\,\zeta(11)-364\,\zeta(13)-1547\,\zeta(15)-2652\,\zeta(17)\ +8398\,\zeta(19)\ \Big]\col\\
\end{aligned}
\end{equation*}
\begin{equation*}
\footnotesize
\begin{aligned}
I_6^{(3)}&=\frac{1}{(4\pi)^{12}}\frac{1}{\varepsilon}\Big[\ \frac{20}{3}\,\zeta(5)-\frac{14}{3}\,\zeta(7)\ \Big]\col\\
I_7^{(3)}&=\frac{1}{(4\pi)^{14}}\frac{1}{\varepsilon}\Big[\ 15\,\zeta(7)-6\,\zeta(9)\ \Big]\col\\
I_8^{(3)}&=\frac{1}{(4\pi)^{16}}\frac{1}{\varepsilon}\Big[\ \frac{63}{4}\,\zeta(7)+\frac{81}{2}\,\zeta(9)-\frac{99}{4}\,\zeta(11)\ \Big]\col\\
I_9^{(3)}&=\frac{1}{(4\pi)^{18}}\frac{1}{\varepsilon}\Big[\ 56\,\zeta(9)+\frac{440}{3}\,\zeta(11)-\frac{286}{3}\,\zeta(13)\ \Big]\col\\
I_{10}^{(3)}&=\frac{1}{(4\pi)^{18}}\frac{1}{\varepsilon}\Big[\
\frac{144}{5}\,\zeta(9)+209\,\zeta(11)+\frac{2431}{5}\,\zeta(13)
-\frac{715}{2}\,\zeta(15)\ \Big]\col\\
I_{11}^{(3)}&=\frac{1}{(4\pi)^{18}}\frac{1}{\varepsilon}\Big[\ 135\,\zeta(11)+819\,\zeta(13)+1638\,\zeta(15)-1326\,\zeta(17)\ \Big]\col\\
& \\
I_8^{(4)}&=\frac{1}{(4\pi)^{16}}\frac{1}{\varepsilon}\Big[\ -\frac{105}{4}\,\zeta(7)-\frac{15}{2}\,\zeta(9)+\frac{165}{4}\,\zeta(11)\ \Big]\col\\
I_9^{(4)}&=\frac{1}{(4\pi)^{18}}\frac{1}{\varepsilon}\Big[\ -56\,\zeta(9)-55\,\zeta(11)+143\,\zeta(13)\ \Big]\col\\
I_{10}^{(4)}&=\frac{1}{(4\pi)^{18}}\frac{1}{\varepsilon}\Big[\
-\frac{336}{5}\,\zeta(9)-231\,\zeta(11)-\frac{429}{5}\,\zeta(13)
+\frac{1001}{2}\,\zeta(15)\ \Big]\col\\
I_{11}^{(4)}&=\frac{1}{(4\pi)^{18}}\frac{1}{\varepsilon}\Big[\ -240\,\zeta(11)-936\,\zeta(13)-182\,\zeta(15)+1768\,\zeta(17)\ \Big]\col\\
& \\
I_{10}^{(5)}&=\frac{1}{(4\pi)^{18}}\frac{1}{\varepsilon}\Big[\
\frac{504}{5}\,\zeta(9)+154\,\zeta(11)-\frac{1144}{5}\,\zeta(13)\ \Big]\col\\
I_{11}^{(5)}&=\frac{1}{(4\pi)^{18}}\frac{1}{\varepsilon}\Big[\ 210\,\zeta(11)+546\,\zeta(13)-637\,\zeta(15)\ \Big]
\pnt
\end{aligned}
\end{equation*}

\normalfont

\normalsize
\bibliographystyle{JHEP}
\bibliography{references-def-L}

\end{fmffile}

\clearpage

\end{document}